\documentclass[journal=jctcce,manuscript=article,layout=traditional]{achemso}

\pdftrailerid{}
\usepackage[T1]{fontenc} \usepackage{mathtools}
\usepackage{graphicx}
\usepackage{lineno}
\usepackage{amsmath}
\usepackage{amssymb}
\usepackage{physics}
\usepackage{mathrsfs}
\usepackage{bm}
\usepackage{threeparttable}
\usepackage{float}
\usepackage{longtable}
\usepackage{multirow}
\usepackage{booktabs}
\usepackage{hyperref}
\usepackage[section]{placeins}
\usepackage{makecell}

\SectionNumbersOn

\makeatletter
\newcommand{\warninput}[1]{\filename@parse{#1}\InputIfFileExists{#1}{}{\message{LaTeX Warning: File `\filename@base.\ifx\filename@ext\relax tex\else\filename@ext\fi' not found on input line \the\inputlineno}}}
\makeatother

\title{Relativistic and Electron Correlation Effects in Static Dipole Polarizabilities for Main-Group Elements}

\author{YingXing Cheng}
\affiliation[University of Stuttgart]{Institute of Applied Analysis and Numerical Simulation Numerical Mathematics for High Performance Computing, University of Stuttgart, Pfaffenwaldring 57, 70569, Stuttgart, Germany}
\email{yingxing.cheng@mathematik.uni-stuttgart.de}

\begin{document}

    \begin{abstract}
        In this study, I compute the static dipole polarizability of main-group elements using the finite-field method combined with relativistic coupled-cluster and configuration interaction simulations.
The computational results closely align with the values recommended in the 2018 table of static dipole polarizabilities of neutral elements [Mol. Phys. 117, 1200 (2019)].
Additionally, I investigate the influence of relativistic effects and electron correlation on atomic dipole polarizabilities.
Specifically, three types of relativistic effects impacting dipole polarizabilities are studied: scalar-relativistic, spin-orbit coupling, and fully relativistic Dirac-Coulomb effects.
The results indicate that scalar-relativistic effects are predominant for atoms in Groups 1--2, with minimal influence from spin-orbit coupling effects.
Conversely, for elements in Groups 13--18, scalar-relativistic effects are less significant, while spin-orbit coupling significantly affects elements starting from the fourth row in Groups 13--14 and from the fifth row in Groups 15--18.
In each category of relativistic effects, the impact of electron correlation is evaluated.
The results show that electron correlation significantly influences dipole polarizability calculations, particularly for Groups 1--2 and 13--14 atoms, but is less significant for Groups 15--18 atoms.
This study provides a comprehensive and consistent dataset of dipole polarizabilities and contributes to a systematic understanding of the roles of relativistic and electron correlation effects in atomic dipole polarizabilities, serving as a valuable reference for future research.
     \end{abstract}

    \newpage
    % \linenumbers

    \section{Introduction}
    \label{sec:introduction}
    The electric dipole polarizability quantifies the response of a system's electron density to an external electric field, leading to induced dipoles.
This property plays a critical role in many applications, such as determining atomic scattering cross sections, refractive indices, dielectric constants, interatomic interactions, and the development of polarizable force fields used in molecular simulations.\cite{Schwerdtfeger2019}
Moreover, an accurate value of the dipole polarizability helps benchmark other methods, such as density-functional theory (DFT).\cite{Bast2008a}

The accurate determination of polarizabilities is essential not only for atomic and molecular physics but also for precision technologies.
For example, in atomic clocks, the leading-order Stark shift and black-body radiation (BBR) shift, which are critical for timekeeping accuracy, depend on precise values of atomic dipole polarizability.\cite{Tang2018,Guo2021}
The BBR shift has contributed to the primary bottleneck of overall uncertainty in strontium (Sr, $Z=38$) and ytterbium (Yb, $Z=70$) clocks for several years at the $10^{-16}$ level.\cite{Ludlow2008,Lemke2009,Ludlow2015}
To further improve the accuracy of atomic clocks, higher precision in the differential static polarizability of the clock states is required.\cite{Ludlow2015}

Although recent studies have reported highly accurate values for several lighter atoms, such as helium (He, $Z=2$)~\cite{Schmidt2007} and neon (Ne, $Z=10$),\cite{Gaiser2010} experimental polarizabilities remain unavailable for most elements.\cite{Schwerdtfeger2019}
Consequently, theoretical calculations are typically employed to obtain these values.
A recent study by Schwerdtfeger and Nagle compiled an updated table of the most accurate calculated and experimental dipole polarizabilities for neutral atoms in the Periodic Table, with nuclear charges ranging from $Z=1$ to $120$, except for livermorium (Lv, $Z=116$).\cite{Schwerdtfeger2019}
However, there remains potential for improvement in the polarizabilities of open-shell elements and heavy atoms that exhibit significant relativistic effects.
This study focuses on all main-group atoms, except for hydrogen (H, $Z=1$), particularly for heavy atoms (i.e., atoms with $Z > 40$).

The determination of accurate atomic polarizabilities requires high-level electron correlation treatment within a relativistic framework, particularly for heavy atoms.
Over the past two decades, researchers have advanced techniques to incorporate spin-orbit coupling (SOC) effects in dipole polarizability calculations.
For closed-shell atoms and open-shell atoms with only one open-shell electron, the coupled cluster theory with single and double substitutions (CCSD) with perturbative triples, CCSD(T), is usually employed within a fully relativistic Dirac-Coulomb framework, where both scalar-relativistic effects and SOC effects can be included.
Due to the SOC effect, atoms in Group 14 are also treated as closed-shell atoms.
However, for open-shell atoms with more than one open-shell electron in Groups 15--17, the coupled cluster theory is not yet well-developed at the fully relativistic level, and an alternative is multi-reference configuration interaction (MRCI).\cite{Schwerdtfeger2019}

Previous studies have thoroughly investigated the polarizabilities considering SOC effects at the CCSD or CCSD(T) level for cesium (Cs, $Z=55$)\cite{Borschevsky2013,Singh2016} and francium (Fr, $Z=87$)\cite{Borschevsky2013,Singh2016a,Aoki2021} in Group 1.
Fleig \textit{et al.} examined the spin-orbit-resolved static polarizabilities of Group 13 atoms, including gallium (Ga, $Z=31$), indium (In, $Z=49$), and thallium (Tl, $Z=81$), using four-component configuration interaction (CI) and coupled cluster methods with ANO-RCC basis sets in the Dirac-Coulomb framework.\cite{Fleig2005}
Borschevsky \textit{et al.} later improved these results using the Fock space coupled cluster (FSCC) with more extensive F{\ae}gri basis sets.\cite{Borschevsky2013}
Pershina \textit{et al.} first reported the polarizabilities of nihonium (Nh, $Z=113$), computed with the FSCC method and a custom F{\ae}gri basis set.\cite{Pershina2008}
Dzuba \textit{et al.} subsequently computed the value for Nh using the SD+CI method in a Dirac framework with Breit interaction and high-order quantum electrodynamics (QED) correlation, agreeing with the previous work.\cite{Dzuba2016}
Fleig and Sadlej investigated elements fluorine (F, $Z=9$), bromine (Br, $Z=35$), iodine (I, $Z=53$), and astatine (At, $Z=85$) with two-component spin-orbit coupling CI calculations,\cite{Fleig2002} while tennessine (Ts, $Z=117$) currently only has an empirically estimated value.\cite{deFarias2017}
No prior study has provided information on the contribution of the SOC effect to Groups 15--16 atoms, with the exception of moscovium (Mc, $Z=115$), which was examined using the SD+CI approach by Dzuba \textit{et al.}\cite{Dzuba2016}
Moreover, errors associated with the recommended values for Groups 15--16 atoms are still significant.\cite{Schwerdtfeger2019}
Currently, no theoretical value exists for Lv.

The main goal of this study is to provide a comprehensive and consistent dataset of dipole polarizabilities for main-group elements.
To achieve this, I employed relativistic single-reference coupled cluster (CC) methods and MRCI methods within the Dirac-Coulomb Hamiltonian framework.
I included more correlated electrons and virtual orbitals/spinors in correlation energy calculations, using larger basis sets than previous theoretical studies.
Additionally, I investigated the impact of electronic correlation and relativistic effects on dipole polarizabilities through a quantitative approach.
It should be noted that the data for light atoms ($Z < 40$) are also computed in this work.
However, their values have been extensively investigated, and improving their theoretical values is beyond the scope of this study.
Readers who are interested in these data can refer to Ref.~\citenum{Schwerdtfeger2019} and its latest version.\cite{Schwerdtfeger2023}

The structure of the remainder of this paper is as follows:
Section~\ref{sec:methods} introduces computational methods, followed by computational details in Section~\ref{sec:details}.
Results and discussions are presented in Section~\ref{sec:results}.
Lastly, a summary is given in Section~\ref{sec:summary}.
Atomic units are used throughout.

    \section{Methods}
    \label{sec:methods}
    \subsection{Relativistic framework}
\label{subsec:relativistic-framework}
In this study, all energies were calculated within a fully relativistic framework using a four-component relativistic formalism.
The relativistic effects can be understood as a combination of SOC effects and scalar-relativistic effects, which account for the contraction or decontraction of the radial electron density.\cite{Yu2015}

\subsubsection{Four-component Dirac-Coulomb Hamiltonian}
An accurate standard for four-component relativistic electronic-structure theory is based on the Dirac-Coulomb-Breit (DCB) Hamiltonian:\cite{Fleig2012}
\begin{align}
    \hat{H}_\text{DCB} = \sum_i \hat{h}_\text{D}(i) + \sum_{i<j} \hat{g}_{ij} + \sum_{A<B}V_{AB},
    \label{eq:H_DCB}
\end{align}
where $i$ and $j$ index electrons, $V_{AB}$ is the interaction between nuclei $A$ and $B$, and $\hat{h}_\text{D}$ is the one-electron Dirac Hamiltonian.
Without external electric fields, it is given by:
\begin{align}
    \hat{h}_\text{D}(i) = c \bm{\alpha} _i \cdot \bm{p}_i + c^2 \beta_i + \sum_{A} V_{iA},
    \label{eq:h_D}
\end{align}
where $c$ is the speed of light and $\bm{p}$ is the momentum operator.
$\bm{\alpha}$ is a vector comprising $\alpha_x$, $\alpha_y$, and $\alpha_z$, which, together with $\beta$, are referred to as the $4 \times 4$ Dirac matrices:
\begin{align}
    \alpha_x =
    \begin{pmatrix}
        0_2      & \sigma_x \\
        \sigma_x & 0_2
    \end{pmatrix},
    \quad
    \alpha_y =
    \begin{pmatrix}
        0_2      & \sigma_y \\
        \sigma_y & 0_2
    \end{pmatrix},
    \quad
    \alpha_z =
    \begin{pmatrix}
        0_2      & \sigma_z \\
        \sigma_z & 0_2
    \end{pmatrix},
    \quad
    \beta =
    \begin{pmatrix}
        I_2      & 0_2      \\
        0_2      & -I_2
    \end{pmatrix},
    \label{eq:alpha}
\end{align}
where $\sigma_x$, $\sigma_y$, and $\sigma_z$ are Pauli spin matrices, and $0_2$ and $I_2$ are $2\times2$ zero and unit matrices, respectively.
The $V_{iA}$ term represents the electron-nuclei interaction on electron $i$ from nucleus $A$.

In the context of two-electron interactions, relativistic corrections up to the second order, specifically the Coulomb-Breit interaction in the Coulomb gauge, are represented by the following equation:\cite{Saue2011}
\begin{align}
    \hat{g}_{ij} = \hat{g}^\text{Coulomb} + \hat{g}^\text{Gaunt} + \hat{g}^\text{gauge}
    = \frac{1}{r_{ij}} - \frac{\bm{\alpha}_i \cdot \bm{\alpha}_j}{r_{ij}}
    + \frac{(\bm{\alpha}_i \cdot \bm{r}_{ij})(\bm{\alpha}_j \cdot \bm{r}_{ij})}{2r_{ij}^3},
    \label{eq:V_ij}
\end{align}
where $\hat{g}^\text{Coulomb}$, $\hat{g}^\text{Gaunt}$, and $\hat{g}^\text{gauge}$ represent the Coulomb interaction, the Gaunt term, and the gauge term, respectively.
$\bm{r}_{i}$ ($\bm{r}_{j}$) is the coordinate of electron $i$ ($j$).
$\bm{r}_{ij} = \bm{r}_i - \bm{r}_j$ and $r_{ij} = \Vert \bm{r}_{ij} \Vert$.
The sum of the Gaunt and gauge terms is also known as the Breit term.
When only the Coulomb interaction is considered, the four-component DCH Hamiltonian reduces to the four-component Dirac-Coulomb (DC) Hamiltonian:
\begin{align}
    \hat{H}_\text{DC} =
    \sum_i \hat{h}_\text{D}(i)
    + \sum_{i<j} \frac{1}{r_{ij}}
    + \sum_{A<B}V_{AB}.
    \label{eq:H_DC_approx}
\end{align}
In this study, all calculations are carried out with the DC Hamiltonian.

\subsubsection{Exact Two-Component Dirac Hamiltonian}
The high computational costs associated with four-component relativistic calculations have prompted the development of methods that use two-component relativistic Hamiltonians.
Examples of these methods include the second-order Douglas-Kroll-Hess (DKH) Hamiltonian~\cite{Douglas1974, Hess1985, Hess1986} and zeroth-order regular approximation (ZORA) Hamiltonians.\cite{Chang1986, vanLenthe1994, vanLenthe1996}
However, these methods are all limited to finite orders.
Ilias \textit{et al.} proposed an infinite-order two-component (IOTC) relativistic Hamiltonian, also known as the exact two-component (X2C) method.\cite{Ilias2007}
In the X2C method, the relativistic effects can be split into a spin-free part and a spin-dependent part, which correspond to scalar-relativistic and the SOC effects, respectively.
In this work, the spin-free X2C method is employed to calculate both non-relativistic and scalar-relativistic properties.
More details about the X2C method can be found in recent reviews.\cite{Saue2011,Saue2020}

\subsection{Relativistic and Correlation Effects}
\label{subsec:electron-correlation}
The Dirac Hartree-Fock (DHF) method was utilized as the uncorrelated method.
In this study, the terms ``orbital'' and ``spinor'' are used to describe the quantum state of an electron without and with considering SOC effects.
Based on these reference states, correlated energies were calculated.
Examples of these methods include the second-order M\text{\o}ller-Plesset (MP2) perturbation theory,\cite{vanStralen2005a} coupled-cluster models CCSD and CCSD(T), where SD stands for single and double excitations, and ``(T)'' represents perturbed triple excitations,\cite{Visscher1996} and the MRCI method with single and double substitutions (MRCISD).\cite{Fleig2003,Fleig2006,Knecht2008,Knecht2010}

In this study, either the four-component (4C) or two-component (2C) Dirac-Coulomb (DC) Hamiltonians are employed to address relativistic effects.
The 4C-DC Hamiltonian, which includes Dirac-Coulomb relativistic effects, while the 2C-DC Hamiltonian, implemented in the X2C method, can be used to incorporate more correlated electrons and active virtual orbitals/spinors.
The notations ``NR-CC'', ``SR-CC'', and ``DC-CC'' are used to refer to CC calculations with non-relativistic, scalar-relativistic, and Dirac-Coulomb relativistic effects, respectively.
MRCISD was also used to study the contribution of Dirac-Coulomb relativistic effect, denoted as ``DC-CI'', for Group 15--17 elements where the single-reference DC-CC method is not available.

In correlated calculations, the occupied and virtual spaces are usually truncated to reduce computational costs.
The orbitals/spinors derived from DHF calculations are then categorized into inner-core, outer-core, valence, and virtual shells.
The inner-core electrons remain inactive during the computation of correlation energy, whereas only the electrons in the outer-core and valence shells are active.
In practice, outer-core shells with energies lower than the energy cutoff and virtual shells with energies exceeding the cutoff are excluded from the calculation.
Thus, one can determine the correlation level of a calculation by considering the number of active electrons and virtual orbitals/spinors.
In general, the inclusion of more active electrons and virtual orbitals/spinors leads to more accurate results.

The notation $\alpha_m^n$ is used to denote dipole polarizabilities computed using a correlated method represented by $m$ and considering a specific relativistic effect denoted by $n$.
For instance, $m$ can be one of the following: DHF, MP2, CCSD, CCSD(T), or MRCISD, while $n$ represents NR, SR, or DC, corresponding to non-relativistic, scalar-relativistic, and Dirac-Coulomb relativistic effects, respectively.
Given a relativistic effect, the corresponding electron-correlation contribution to polarizabilities is defined as:
\begin{align}
    \Delta \alpha_c^n := \alpha_{\text{CCSD(T)}}^n - \alpha_{\text{DHF}}^n
    \label{eq:alpha_corr}
\end{align}
where $\Delta \alpha_c^n$ represents the contribution of electron correlation specific to the relativistic effect denoted by $n$.

Next, the impact of relativistic effects on polarizabilities are estimated.
The scalar-relativistic effect on polarizabilities is defined as follows:
\begin{align}
    \Delta \alpha_r^\text{SR} := \alpha_\text{CCSD(T)}^\text{SR} - \alpha_\text{CCSD(T)}^\text{NR}
    \label{eq:rel_effects_sr}
\end{align}
where $\alpha_\text{CCSD(T)}^\text{SR}$ and $\alpha_\text{CCSD(T)}^\text{NR}$ are scalar-relativistic and non-relativistic values computed by the CCSD(T) method with the same correlation level.

In addition, the SOC effect on dipole polarizabilities can be then defined as:\cite{Fleig2005,Yu2015}
\begin{align}
\Delta \alpha_r^\text{SOC} =
\begin{cases}
    \alpha_\text{CCSD(T)}^\text{DC} - \alpha_\text{CCSD(T)}^\text{SR} & \text{if DC-CC availiable}  \\
    \alpha_\text{MRCISD}^\text{DC} - \alpha_\text{CCSD(T)}^\text{SR} & \text{if DC-CC not availiable}
\end{cases}
    \label{eq:rel_effects_so}
\end{align}
where $\alpha_\text{CCSD(T)/MRCISD}^\text{DC}$ denotes the DC relativistic polarizabilities computed by the CCSD(T) method, or MRCISD values.
While an identical number of active electrons and virtual orbitals/spinors is used for both DC and SR calculations, directly comparing MRCISD values to CCSD(T) data remains challenging due to the contribution from higher-order excitations missed in the MRCISD method.
Therefore, Eq.~\eqref{eq:rel_effects_so} represents both the SOC effects and high excitations when MRCISD values are utilized.
To exclusively study the SOC effects, one should compare DC-CI values with MRCISD values that incorporate SR effects, which is beyond the scope of this work.
Moreover, the correlation level used in scalar-relativistic calculations might differ between Eqs.~\eqref{eq:rel_effects_sr} and~\eqref{eq:rel_effects_so}.
This is because DC calculations at the same correlation level are much more costly than their SR counterparts.
Consequently, when the correlation levels differ, notations SR$_n$ and SR$_d$ are used to refer to the values defined in Eqs.~\eqref{eq:rel_effects_sr} and~\eqref{eq:rel_effects_so}, respectively.

The contribution of full-relativistic DC effects, denoted as $\Delta \alpha_r^\text{DC}$, can be calculated from the sum of contributions from scalar-relativistic effects and the SOC effect:
\begin{align}
    \Delta \alpha_r^\text{DC} = \Delta \alpha_r^\text{SR} + \Delta \alpha_r^\text{SOC}.
    \label{eq:rel_effects_dc}
\end{align}

\subsection{Open-shell methods}\label{subsec:open-shell-methods}
For the spin-dependent coupling case ($jj$ representation), the averaged polarizability $\bar{\alpha}^J$ for the ground state is determined by:\cite{Yu2015}
\begin{align}
    \bar{\alpha}^J = \frac{1}{2J + 1} \sum_{M_J} \alpha(J, M_J)
    \label{eq:Q_J}
\end{align}
where $\alpha(J, M_J)$ denotes the polarizabilities for each $J, M_J$ component.
However, for spin-free cases ($LS$ representation), the equation is analogous to Eq.~\eqref{eq:Q_J} with $J$ replaced by $L$:\cite{Yu2015}
\begin{align}
    \bar{\alpha}^L = \frac{1}{2L + 1} \sum_{M_L} \alpha(L, M_L)
    \label{eq:Q_L}
\end{align}

In this work, only $p$-block open-shell elements are studied.
The atomic $p$ spinors in the relativistic case are linear combinations of the real $p$ orbitals in the non-relativistic case:\cite{Fleig2002}
\begin{align}
    \phi \left( \frac{3}{2},  \frac{3}{2} \right)
    &= \sqrt{\frac{1}{2}} (p_x + i p_y)|\uparrow \rangle
    &
    \phi \left( \frac{3}{2}, -\frac{3}{2} \right)
    &= \sqrt{\frac{1}{2}} (p_x - i p_y)|\downarrow \rangle
    \nonumber
    \\
    \phi \left( \frac{3}{2},  \frac{1}{2} \right)
    &= \sqrt{\frac{2}{3}} p_z |\uparrow \rangle + \sqrt{\frac{1}{6}}(p_x + ip_y)|\downarrow \rangle
    &
    \phi \left( \frac{3}{2}, -\frac{1}{2} \right)
    &= \sqrt{\frac{2}{3}} p_z |\downarrow \rangle + \sqrt{\frac{1}{6}}(p_x - ip_y)|\uparrow \rangle
    \nonumber
    \\
    \phi \left( \frac{1}{2},  \frac{1}{2} \right)
    &= - \sqrt{\frac{1}{3}} p_z |\uparrow   \rangle  + \sqrt{\frac{1}{3}}(p_x + ip_y)|\downarrow \rangle
    &
    \phi \left( \frac{1}{2}, -\frac{1}{2} \right)
    &=   \sqrt{\frac{1}{3}} p_z |\downarrow \rangle  - \sqrt{\frac{1}{3}}(p_x - ip_y)|\uparrow   \rangle,
    \label{eq:p_spinors}
\end{align}
where, $|\uparrow \rangle$ and $|\downarrow \rangle$ are spin-up and spin-down states in non-relativistic case, respectively.
The spinors $\phi \left( \frac{1}{2}, \pm \frac{1}{2} \right)$ have the lowest energy, whereas the spinors $\phi \left( \frac{3}{2}, \pm \frac{3}{2} \right)$ and spinors $\phi \left( \frac{3}{2}, \pm \frac{1}{2} \right)$ are degenerate without perturbation.

\subsubsection{Group 13 elements}
The spin-orbit components of Group 13 atomic states can be represented as the $p^1$ states using atomic spinors $\phi(j,m_j)$ due to the presence of only one electron in $p$-spinor.
In the ground state, only the spinors $\phi \left( \frac{1}{2}, \pm \frac{1}{2} \right)$ are occupied.
When an external perturbation, such as an electric field along the $z$ axis, is applied, the two degenerate spinors $\phi \left( \frac{3}{2}, \pm \frac{3}{2} \right)$ and spinors $\phi \left( \frac{3}{2}, \pm \frac{1}{2} \right)$ split, resulting in a non-degenerate state.
The energy shift is dependent on the $p_z$-component and generally, spinors with a larger $p_z$ contribution will experience a larger energy shift than those with a smaller contribution.\cite{Fleig2002}
However, as weak electric fields are typically applied, the energy shift is much smaller in absolute value than the spin-orbit splitting.
As a result, the $J=\frac{3}{2}, M_J=\pm \frac{3}{2}$ states will have a smaller dipole polarizability than the $J=\frac{3}{2}, M_J=\pm \frac{1}{2}$ states.

Based on the analysis in Ref.~\citenum{Fleig2005}, the value $\bar{\alpha}_{J=3/2} - \alpha_{J=1/2}$ can be interpreted as the effect of both SOC and the contraction (decontraction) of the $p_{1/2}$ ($p_{3/2}$) spinor in total.
The SOC effect can also be determined by comparing the $M_L$-averaged polarizability, calculated using SR-CC methods, to the value of the $J=\frac{1}{2}$ ground state computed using the DC-CC method.
Moreover, the DC-CI method allows for the computation of excited-state polarizabilities, facilitating the study of spin-orbit resolved polarizabilities for these excited state components.

\subsubsection{Group 14 elements}
As previously discussed, the $J=\frac{1}{2}, M_J=\frac{1}{2}$ spinors have a lower energy than the $\phi\left(\frac{3}{2}, m_j\right)$ spinors, resulting in the full occupation of the $\phi\left(\frac{1}{2},m_j\right)$ spinors.
This means that Group 14 atoms can be considered as closed-shell atoms due to the SOC effect.

\subsubsection{Group 15 elements}
The DC-CI calculations were performed to investigate the electron-correlation and the SOC effect on polarizabilities.
Additionally, the polarizabilities of different $M_J$ sub-states were studied.
In practice, DC-CI eigenvectors are often represented by string occupation and the spin-projection value $M_K$, which is an integer multiple of $\frac{1}{2}$ for a given determinant.\cite{Fleig2002}
As a result, the different components of ${^4}P_{3/2}$ are not pure states, and may be combined with high $M_K$ states that have the same $M_J$.
To address this, the method proposed by Fleig \textit{et al.}~\cite{Fleig2002} was used to approximately specify the $M_J$ state.
However, since the primitive determinant is determined by $M_J=\pm 3/2$ and $M_J=\pm 1/2$ for $M_K = \frac{1}{2}$, the polarizabilities of these states are thus approximately considered to represent ground-state values.
In this work, the states $J=\frac{3}{2}, M_J=\pm \frac{1}{2}$ refer to the states with lower energy compared to the states $J=\frac{3}{2}, M_J=\pm \frac{3}{2}$ when an external electric field is applied along the $z$-axis.

\subsubsection{Group 16 elements}
The polarizabilities of different $M_J$ components of the ${^3}P_2$ states were determined using a similar procedure to that used for Group 15 elements.
The DC-CI method was used to calculate the polarizabilities for states $M_J=0$, $M_J=\pm1$ and $M_J=\pm2$, and the $M_J$-averaged polarizability is approximated as the ground-state value.
In this work, the states $J=2, M_J=\pm 1$ refer to the states with lower energy compared to the states $J=2, M_J=\pm 2$ when an external electric field is applied along the $z$-axis.

\subsubsection{Group 17 elements}
The spin-orbit components of the ground states of halogen atoms can be represented as the inverse of the $p^5$ hole states by using atomic spinors $\phi(j,m_j)$ as there is only one hole in the $p$-spinor for these atoms.\cite{Fleig2002}
The polarizabilities of different components of the ${^2}P_{3/2}$ states were determined using a similar procedure to that used for Group 15 elements.
The DC-CI method was used to calculate the polarizabilities for states $M_J=\pm3/2$ and $M_J=\pm1/2$, and the $M_J$-averaged polarizability is approximated as the ground-state value.
Moreover, the computational values for states ${^2}P_{1/2}$ are also computed.

\subsection{Finite-field methods}
\label{subsec:finite-field-methods}
The static dipole polarizabilities presented in this work were computed using finite-field methods.~\cite{Das1998}.
The energy of an atom in a homogeneous electric field of strength $F_z$ along the $z$-axis is represented as follows:\cite{Das1998}
\begin{align}
    E(F_z) = E_0 -\frac{1}{2}\alpha F_z^2 - \frac{1}{4!}\gamma  F_z^4 - \frac{1}{6!}\gamma_2 F_z^6 - \cdots
    \label{eq:ff_d}
\end{align}
where $E_0$ denotes the field-free energy, while $\alpha$, $\gamma$, and $\gamma_2$ correspond to the dipole polarizability, the dipole hyperpolarizability, and the second dipole hyperpolarizability, respectively.
For the purpose of this study, the expansion is truncated at the third term, allowing one to calculate the $\alpha$ and $\gamma$:
\begin{align}
    E(F_z) \approx E_0 -\frac{1}{2}\alpha F_z^2 - \frac{1}{4!}\gamma  F_z^4.
    \label{eq:ff_d_2}
\end{align}
The least-squares procedure is used instead of numerical differentiation for reasons discussed in Ref.~\citenum{Kassimi1994}.
In practice, this procedure could lead to an unreasonable $\gamma$, including extremely positive or negative values.
In these cases, an expansion only including $\alpha$ is also used:
\begin{align}
    E(F_z) \approx E_0 -\frac{1}{2}\alpha F_z^2.
    \label{eq:ff_d_1}
\end{align}
The results obtained from Eq.~\eqref{eq:ff_d_2}, denoted as $\alpha(2)$, should always be used instead of the values obtained from Eq.~\eqref{eq:ff_d_1}, denoted as $\alpha(1)$, except in cases where all the following conditions are satisfied:
\begin{enumerate}
    \item The relative difference between $\alpha(1)$ and $\alpha(2)$, denoted as $\tau$, is greater than 0.6\%, i.e.,
        \begin{align}
            \tau = \frac{|\alpha(2) - \alpha(1)|}{\alpha(2)} \times 100\% > 0.6\%.
            \label{eq:tau}
        \end{align}

    \item An invalid $\gamma$ is obtained from Eq.~\eqref{eq:ff_d_2}.
        For simplicity, in this work, an invalid $\gamma$ is defined as when at least one of the following conditions is satisfied:
        \begin{enumerate}
            \item For CCSD(T) methods,
            \begin{align}
                \gamma_\text{CCSD(T)} & < 0, &
                \gamma_\text{CCSD} & < 0, &
                \epsilon =\frac{\max(\gamma_\text{CCSD(T)}, \gamma_\text{CCSD})}{\min(\gamma_\text{CCSD(T)}, \gamma_\text{CCSD})} & > 10,
            \end{align}
            where $\gamma_\text{CCSD(T)}$ and $\gamma_\text{CCSD}$ are the $\gamma$ values obtained by the CCSD(T) and CCSD methods, respectively.
            Here, $\max(a, b)$ represents the maximum value between $a$ and $b$, and $\min(a, b)$ represents the minimum value between $a$ and $b$.
            $\epsilon$ represents the ratio of the maximum to the minimum of these $\gamma$ values, indicating a significant difference between the $\gamma$ values from the two methods when $\epsilon > 10$.
        \item For MRCISD methods:
            \begin{align}
                \gamma_\text{MRCISD} & < 0,
            \end{align}
        \end{enumerate}
            where $\gamma_\text{MRCISD}$ refers to the $\gamma$ value obtained by the MRCISD method.
\end{enumerate}

\subsection{Uncertainty estimation}
\label{subsec:uncertainty}

The uncertainty estimation is analyzed empirically based on the composite scheme, which has been widely employed in previous studies to obtain the final value of an atomic property, denoted as $P_\text{final}$.\cite{Kallay2011,Yu2015,Irikura2021}
It should be noted that the $P_\text{final}$ values are not provided in this work due to their empirical nature and are only used for uncertainty estimation empirically.

\subsubsection{Composite scheme of the CC method}

Generally, in the CC calculations,

\begin{align}
    P_\text{final} = P_\text{CCSD} + \Delta P_\text{basis} + \Delta P_\text{(T)} \text{ or } \Delta P_\text{T} + \Delta P_\text{Q} + \Delta P_\text{core} + \Delta P_\text{vir} + \Delta P_\text{fitting} + \Delta P_\text{others}
\end{align}

with

\begin{align}
    \Delta P_\text{(T)} &= P_\text{CCSD(T)} - P_\text{CCSD} \\
    \Delta P_\text{T} &= P_\text{CCSDT} - P_\text{CCSD} \\
    \Delta P_\text{Q} &= P_\text{CCSDTQ} - P_\text{CCSDT}
\end{align}

where $P_\text{CCSD}$, $P_\text{CCSD(T)}$, $P_\text{CCSDT}$, and $P_\text{CCSDTQ}$ are the values calculated for the property using CCSD, CCSD(T), CCSDT (CC singles, doubles, and triples), and CCSDTQ (CC singles, doubles, triples, and quadruples), respectively.
$\Delta P_\text{basis}$ is the error due to the use of the finite-size basis functions for computing $P_\text{CCSD}$ or $P_\text{CCSD(T)}$.\cite{Guo2021,Irikura2021}
$\Delta P_\text{core}$ is the correction due to the more outer-core shells, while $\Delta P_\text{vir}$ is the possible error due to the neglected contributions from the frozen high-lying virtual orbitals/spinors.\cite{Guo2021}
$P_\text{CCSD}$ or $P_\text{CCSD(T)}$ is the value for property $P$ using the most accurate CCSD or CCSD(T) method in this work.
$\Delta P_\text{fitting}$ is the numerical error resulting from the fitting procedure using Eq.~\eqref{eq:ff_d_2} or Eq.~\eqref{eq:ff_d_1}.
For completeness, $\Delta P_\text{others}$ corresponds to all other corrections, e.g., the difference between $\Delta P_\text{(T)}$ and $\Delta P_\text{T}$ when $\Delta P_\text{(T)}$ is used, or other relativistic effects, including Gaunt, Breit, or QED interactions.

It should be noted that the most accurate calculation in CC used in this work is CCSD(T).
When CCSDT and CCSDTQ values are available in the literature, $\Delta P_\text{T}$ and $\Delta P_\text{Q}$ are employed; otherwise, $\Delta P_\text{(T)}$ is used, and $\Delta P_\text{Q}$ is neglected with the inclusion of a larger empirical error.
Also, $\Delta P_\text{others}$ is included when its valid value is available in the literature.

\subsubsection{Composite scheme of the MRCI method}

In the CI calculation,

\begin{align}
    P_\text{final} = P_\text{MRCISD} + \Delta P_\text{basis} + \Delta P_\text{T} + \Delta P_\text{Q} + \Delta P_\text{core} + \Delta P_\text{vir} + \Delta P_\text{fitting} + \Delta P_\text{others}
\end{align}

with

\begin{align}
    \Delta P_\text{T} &= P_\text{MRCISDT} - P_\text{MRCISD} \\
    \Delta P_\text{Q} &= P_\text{MRCISDTQ} - P_\text{MRCISDT}
\end{align}

where $P_\text{MRCISD}$, $P_\text{MRCISDT}$, and $P_\text{MRCISDTQ}$ are the values calculated for the property using MRCISD, MRCISDT (MRCI singles, doubles, and triples), and MRCISDTQ (MRCI singles, doubles, triples, and quadruples), respectively.
$\Delta P_\text{basis}$ is the error due to the use of the finite-size basis functions for computing $P_\text{MRCISD}$.\cite{Guo2021,Irikura2021}
$\Delta P_\text{core}$ is the correction due to the more outer-core shells, while $\Delta P_\text{vir}$ is the possible error due to the neglected contributions from the frozen high-lying virtual spinors.\cite{Guo2021}
$\Delta P_\text{fitting}$ is the numerical error resulting from the fitting procedure using Eq.~\eqref{eq:ff_d_2} or Eq.~\eqref{eq:ff_d_1}.
Similarly, $\Delta P_\text{others}$ corresponds to all other corrections as in CC calculations.

It should be noted that the highest-order calculation in CI is MRCISD in this work.
When MRCISDT and MRCISDTQ values are available in the literature, $\Delta P_\text{T}$ and $\Delta P_\text{Q}$ are employed; otherwise, they are neglected with a larger empirical error used.

\subsubsection{Errors estimation}

Based on the composite schemes, the possible error sources causing the uncertainty in $P_\text{final}$ in this work are similar to those proposed in Refs.~\citenum{Yu2015,Guo2021,Irikura2021}:

\begin{enumerate}
    \item The first one is due to the finite basis sets used for computing $P_\text{CCSD}$/$P_\text{CCSD(T)}$ and $P_\text{MRCISD}$, i.e., $\Delta P_\text{basis}$.
    Since the errors converge quickly, the errors in these values are taken as half of the differences of $P$ values obtained by ``d-aug-dyall.v3z'' and ``d-aug-dyall.v4z'', corresponding to the two most diffuse exponents for the respective atomic shells in the original basis sets ``dyall.v3z'' and ``dyall.v4z'', respectively.\cite{Yu2015}
    In this work, all most accurate calculations of $p$-block elements are carried out using ``dyall.acv4z'', where one more diffuse but more core basis functions are added to the original ``dyall.v4z'' basis set.
    See Ref.~\citenum{dirac_basis} for more details.
    This encourages the belief that the results in this work are also closer to the complete basis sets.
    For simplicity, the same empirical estimate method defined in Ref.~\citenum{Yu2015} is used when results with additional basis functions are available in this work.
    Otherwise, a conservative choice is to include the full difference between this work and the most accurate results obtained from the literature as the error.
    \item The second error is from $\Delta P_\text{(T)}$, $\Delta P_\text{T}$, or $\Delta P_\text{Q}$.
    Half of them is also taken into account as the error as in Ref.~\citenum{Yu2015}.
    \item The third one is from $\Delta P_\text{core}$ due to the number of outer-core shells included in the calculations.
    \item The fourth one is from $\Delta P_\text{vir}$ due to neglecting contributions from the outer-core shells.
    \item The next one is from $\Delta_\text{fitting}$ due to the numerical fitting procedure.
Specifically, the standard errors of the regression coefficients are evaluated using the residuals from the least-square solutions,~\cite{Williams2016} which has been implemented in Ref.~\citenum{pydirac}.
    \item The last one is the error due to $\Delta P_\text{others}$.
    It is usually computationally demanding to compute them systematically in practice.
    For simplicity, it is only included when available in the previous literature.
\end{enumerate}

The uncertainty of the final $P_\text{final}$ values is determined by taking the quadrature of the $\Delta P_\text{basis}$, $\Delta P_\text{(T)}$, $\Delta P_\text{Q}$, $\Delta P_\text{core}$, $\Delta P_\text{vir}$, $\Delta P_\text{fitting}$, and $\Delta P_\text{others}$ contributions.

    \section{Computational Details}
    \label{sec:details}
    In this section, the computational details are introduced, including the computational basis sets used and the selected correlation level.

\subsection{Basis sets}\label{subsec:basis_sets}
In this work, I mainly used the Dyall quadruple-$\zeta$ family uncontracted basis sets.\cite{Dyall2002,Dyall2004,Dyall2006,Dyall2007,Dyall2009,Dyall2010,Dyall2011,Dyall2016}
For Group 1--2 atoms, I used ANO-RCC basis set~\cite{Roos2004,Roos2005} instead of the Dyall basis set.
The original ANO-RCC basis sets are augmented with additional functions, extending each type of function in an even-tempered manner.
The exponential coefficients of the additional augmented functions were determined using the equation $\zeta_{N+1} = \zeta_N^2 /\zeta_{N-1}$, where $\zeta_N$ and $\zeta_{N-1}$ represent the smallest exponents for each atomic shell in the default basis sets.\cite{Yu2015}
These new basis sets are labeled as ``s-aug-ANO-RCC'' and ``d-aug-ANO-RCC'' for single and double augmentations, respectively.
After test calculations, the ``s-aug-ANO-RCC'' basis set was selected for Group 1 elements.
For Group 2 elements, the ``s-aug-dyall.cv4z'' basis set was used for Sr, Ba, and Ra, while the ``d-aug-dyall.cv4z'' basis set was utilized for Be, Mg, and Ca.
In addition, ``dyall.cv4z'' and ``d-aug-dyall.cv4z'' are also used in test calculations to assess the error due to finite basis sets, which will be discussed later.
For $p$-block elements, I used the ``dyall.acv4z'' basis set, which includes one additional diffuse and more core basis functions compared to the original ``dyall.v4z'' basis set.

\subsection{Correlation level}
\label{subsec:correlation_level}
All calculations are performed using the DIRAC18 package.\cite{DIRAC18}
In general, orbitals within an energy range of -20 to 25 a.u.\ are correlated in CC and CI calculations.
The convergence criterion for the CC method was set at $10^{-10}$.
In SR-CC calculations, any deviation between SR$_n$ and SR$_d$ values was observed to be less than 1\% in this work.
For the MRCISD calculations, the energy criterion for convergence was set to $10^{-13}$.
Further computational details for $p$-block open-shell elements are provided in the following subsections.

\subsubsection{Group 13 elements}
DHF orbitals, used for DC-CC calculations, are optimized by averaging over various $np$ occupations with one open-shell electron.
The open shells are restricted to the $np_{1/2}$ spinors, leading to a single determinant reference as detailed in previous work~\cite{Fleig2005}.
Additionally, DC-CI calculations were employed to obtain excited-state components, with DHF orbitals constructed from averaging over all $np_{1/2}$ and $np_{3/2}$ spinors.

\subsubsection{Group 14 elements}
In DC calculations, Group 14 elements are typically considered as closed-shell systems.
Consequently, the 2C-DC Hamiltonian was used in all DC-CC calculations for Group 14 elements, allowing for the inclusion of more correlated electrons.
However, this approach is not applicable to the carbon atom, as the $\phi\left(\frac{1}{2},m_j\right)$ spinor occupied state does not constitute a suitable reference wave function, which can lead to convergence issues.
Therefore, the DC calculation for the carbon atom was conducted using the 4C-DC-CC method.

\subsubsection{Group 15--17 elements}
Only DC-CI calculations were performed to explore the effects of electron correlation and the contribution of the SOC effect on polarizabilities.
The multiple-reference wavefunction is used in the DHF method of DC-CI calculation, while the single-reference wavefunction is used in DHF of SR-CC calculation.
Therefore, DHF energies obtained in DC-CI calculations are not employed here to compute properties related to $\alpha_{\text{DHF}}$.
As a result, $\Delta \alpha_c^\text{SOC}$ is not available for Group 15--17 elements.

\subsection{Finite-field method}
\label{subsec:finite-field-method}
Electric fields with strengths of 0.000, 0.0005, 0.001, 0.002, and 0.005 a.u. are applied to each element in the calculation of dipole polarizabilities, with the exception of Group 1 elements.
For these elements, a new set of uniform electric fields with strengths of 0.000, 0.0005, 0.001, 0.0015, and 0.002 a.u. is applied due to their large polarizabilities.
For scalar-relativistic CC methods, a larger range of finite electric fields, ranging from 0.000 to 0.020 a.u. in increments of 0.001, is incorporated to obtain more accurate results for $p$-block elements.
The specified number of electric fields for each atom can be found in the Supporting Information.
The energies are then fitted to Eq.~\eqref{eq:ff_d} using a least-squares method to obtain the dipole polarizability.

    \section{Results}
    \label{sec:results}
    In this section, I present the computational values for main-group elements and discuss the impact of relativistic effects and electron correlation on dipole polarizabilities.
All results related to $\alpha$ obtained by fitting Eq.~\eqref{eq:ff_d_2} are provided in Tables S1-S8 in the Supporting Information.
Results for $\alpha$ obtained by fitting Eq.~\eqref{eq:ff_d_1} can be found in Tables S9-S16 in the Supporting Information.
It was observed that for the most accurate DC values, only aluminum (Al, $Z=13$) should be updated using Eq.~\eqref{eq:ff_d_1} when $\tau > 0.6\%$.
Additionally, Al and oganesson (Og, $Z=118$) should be updated using Eq.~\eqref{eq:ff_d_1} when $\tau > 0.5\%$.
For $\tau > 0.1\%$, the DC values of potassium (K, $Z=19$), rubidium (Rb, $Z=37$), Al, Nh, and Og should be computed using Eq.~\eqref{eq:ff_d_1}.
Therefore, unless otherwise specified, the results derived from Eq.~\eqref{eq:ff_d_2} are consistently utilized in the following discussions.

\autoref{tab:dipole_summary} summarizes the CCSD(T) and MRCISD results for main-group elements, considering non-relativistic, scalar-relativistic, and full Dirac-Coulomb relativistic effects, and compares them with the recommended values from Ref.~\citenum{Schwerdtfeger2019}.
Additionally, component-resolved results are presented for the ${^2}P_{3/2}$ state of Group 13 elements, the ${^4}S_{3/2}$ state of Group 15 elements, the ${^3}P_2$ state of Group 16 elements, and the ${^2}P_{1/2}$ state of Group 17 elements.
The corresponding results derived from Eq.~\eqref{eq:ff_d_1} are available in Table S17 in the Supporting Information.

Next, only a summary of the comparison with the recommended values from Ref.~\citenum{Schwerdtfeger2019} is presented.
A detailed comparison with other computational and experimental values from the literature can also be found in the Supporting Information.
Moreover, the final uncertainty for each atom is provided in \autoref{tab:dipole_summary}, with a detailed uncertainty estimation for each atom available in the Supporting Information.

{
\tiny
\begin{longtable}{llllllllll}
\caption{Static dipole polarizabilities (in a.u.) with non-relativistic (NR), scalar-relativistic (SR), and full Dirac-Coulomb (DC) relativistic effects for main-group elements. The error due to the numerical fitting procedure ($\Delta P_\text{fitting}$) is included as the error bar. The recommended values (Rec.), including the uncertainty estimation as the error bar, are also listed and compared to the counterparts from Ref.~\citenum{Schwerdtfeger2019}.}
\label{tab:dipole_summary} \\
\toprule
\multirow{2}{*}{$\hat{H}$} & \multirow{2}{*}{State} & \multirow{2}{*}{Method} & \multicolumn{7}{c}{\textbf{Period}} \\
\cmidrule(lr){4-10}
 &  &         &              1   &  2  &                 3 &                  4 &                 5 &                 6 &                 7 \\
\midrule
\endfirsthead

\multicolumn{10}{c}{{\bfseries \tablename\ \thetable{} -- continued from previous page}} \\
\toprule
\multirow{2}{*}{$\hat{H}$} & \multirow{2}{*}{State} & \multirow{2}{*}{Method} & \multicolumn{7}{c}{\textbf{Period}} \\
\cmidrule(lr){4-10}
&  &         &              1   &  2  &                 3 &                  4 &                 5 &                 6 &                 7 \\
\midrule
\endhead

\midrule \multicolumn{10}{r}{{Continued on next page}} \\
\endfoot

\bottomrule
\endlastfoot
\multicolumn{3}{c}{\textbf{Group 1}}   &          &  Li &                 Na &                  K &                 Rb &                 Cs &                 Fr \\
\midrule
NR & $^2S$ & CCSD(T) &           & $164.34$ &           $163.98$ &           $295.56$ &  $344.65 \pm 0.01$ &  $474.79 \pm 0.05$ &  $541.62 \pm 0.04$ \\
SR & $^2S$ & CCSD(T) &           & $164.29$ &           $162.99$ &           $289.96$ &  $319.34 \pm 0.01$ &  $397.40 \pm 0.02$ &  $319.36 \pm 0.01$ \\
DC & $^2S_{1/2}$ & CCSD(T) &     & $164.31 \pm 0.01$ &  $163.27 \pm 0.09$ &  $290.64 \pm 0.24$ &  $320.08 \pm 0.29$ &  $398.24 \pm 0.41$ &  $317.40 \pm 0.57$ \\
Rec. & $--$ & $--$ &             & $164.31 \pm 0.20$ &  $163.27 \pm 0.81$ &  $289.40 \pm 2.13$ &  $320.08 \pm 0.83$ &  $398.24 \pm 2.71$ &  $317.40 \pm 0.99$ \\
Ref.~\citenum{Schwerdtfeger2019}& $--$ & $--$ &  &           $164.11$ &      $162.7\pm0.5$ &      $289.7\pm0.3$ &      $319.8\pm0.3$ &      $400.9\pm0.7$ &      $317.8\pm2.4$ \\
\midrule
 \multicolumn{3}{c}{\textbf{Group 2}}              &      &                Be &                Mg &                 Ca &                 Sr &                 Ba &                 Ra \\
\midrule
NR & $^1S$ & CCSD(T) &         &    $37.80$ &           $71.54$ &           $159.90$ &  $209.64 \pm 0.01$ &  $316.58 \pm 0.01$ &  $378.82 \pm 0.01$ \\
SR & $^1S$ & CCSD(T) &         &    $37.79$ &           $71.20$ &           $157.61$ &           $197.82$ &           $273.96$ &  $247.69 \pm 0.01$ \\
DC & $^1S_0$ & CCSD(T) &       &      $37.78$ &           $71.19$ &           $157.55$ &  $197.67 \pm 0.01$ &           $273.51$ &  $245.68 \pm 0.01$ \\
Rec. & $--$ & $--$ &  &  $37.78 \pm 0.10$ &  $71.19 \pm 0.32$ &  $157.55 \pm 1.42$ &  $197.67 \pm 2.31$ &  $273.51 \pm 4.15$ &  $245.68 \pm 4.91$ \\
Ref.~\citenum{Schwerdtfeger2019} &$--$ &  $--$ &  &    $37.74\pm0.03$ &      $71.2\pm0.4$ &        $160.8\pm4$ &      $197.2\pm0.2$ &       $272.9\pm10$ &        $246.0\pm4$ \\
\midrule
\multicolumn{3}{c}{\textbf{Group 13}}          &      &                 B &                Al &                Ga &                In &                Tl &                 Nh \\
\midrule
NR & $^2P$ & CCSD(T) &           &  $20.41$ &  $55.50 \pm 0.47$ &  $52.76 \pm 0.05$ &  $69.25 \pm 0.05$ &  $70.61 \pm 1.10$ &   $88.88 \pm 0.70$ \\
SR$_n$ & $^2P$ & CCSD(T) &       &  $20.40 \pm 0.01$ &  $55.54 \pm 0.43$ &  $52.30 \pm 0.13$ &  $66.91 \pm 0.20$ &  $64.37 \pm 0.94$ &   $81.62 \pm 0.11$ \\
SR$_d$ & $^2P$ & CCSD(T) &       &  $20.40 \pm 0.01$ &  $55.54 \pm 0.43$ &  $52.00 \pm 0.09$ &  $66.98 \pm 0.07$ &  $64.53 \pm 1.01$ &   $81.96 \pm 0.11$ \\
DC & $^2P_{1/2}$ & CCSD(T) &     &          $--$ &  $57.51 \pm 1.75$ &  $51.08 \pm 0.32$ &           $61.92$ &  $51.26 \pm 0.13$ &   $32.89 \pm 0.18$ \\
                                 &      & MRCISD &         &    $20.35$ &  $57.61 \pm 0.02$ &           $50.66$ &           $62.87$ &           $51.26$ &            $28.21$ \\
                                 & $^2P_{3/2}$ & MRCISD &        &     $20.38$ &  $58.07 \pm 0.01$ &           $52.86$ &           $70.55$ &           $81.71$ &  $129.98 \pm 0.01$ \\
                                 & $^2P_{3/2}, M_J=1/2$ & MRCISD &        &     $22.31$ &  $66.81 \pm 0.01$ &           $63.64$ &           $84.75$ &          $106.41$ &  $188.76 \pm 0.01$ \\
                                 & $^2P_{3/2}, M_J=3/2$ & MRCISD &        &     $18.46$ &           $49.33$ &           $42.08$ &           $56.36$ &           $57.01$ &            $71.21$ \\
Rec. & $--$ & $--$ &  &  $20.40 \pm 0.09$ &  $57.51 \pm 2.47$ &  $51.08 \pm 1.19$ &  $61.92 \pm 1.92$ &  $51.26 \pm 1.04$ &   $32.89 \pm 1.60$ \\
Ref.~\citenum{Schwerdtfeger2019} & $--$ & $--$ &     &     $20.5\pm1$ &      $57.8\pm1.0$ &        $50.3\pm3$ &          $65\pm4$ &        $50.0\pm2$ &         $29.2\pm2$ \\
\midrule
\multicolumn{3}{c}{\textbf{Group 14}}           &      &                 C &                Si &                Ge &                Sn &                Pb &                Fl \\
\midrule
NR & $^3P$ & CCSD(T) &           &  $11.66$ &  $37.07 \pm 0.01$ &  $39.70 \pm 0.16$ &  $55.48 \pm 0.15$ &  $62.20 \pm 0.09$ &  $77.92 \pm 0.13$ \\
SR$_n$ & $^3P$ & CCSD(T) &       &      $11.67$ &  $37.10 \pm 0.01$ &  $39.53 \pm 0.07$ &  $54.18 \pm 0.13$ &  $58.49 \pm 0.10$ &  $72.54 \pm 0.15$ \\
SR$_d$ & $^3P$ & CCSD(T) &       &      $11.67$ &  $37.10 \pm 0.01$ &  $39.70 \pm 0.03$ &  $54.30 \pm 0.04$ &  $58.49 \pm 0.10$ &  $72.54 \pm 0.15$ \\
DC & $^3P_0$ & CCSD(T) &         &    $11.73$ &  $37.00 \pm 0.01$ &  $39.19 \pm 0.02$ &           $51.96$ &           $46.53$ &           $30.77$ \\
Rec. & $--$ & $--$ &  &  $11.73 \pm 0.48$ &  $37.00 \pm 0.63$ &  $39.19 \pm 0.74$ &  $51.96 \pm 1.01$ &  $46.53 \pm 0.89$ &  $30.77 \pm 0.90$ \\
Ref.~\citenum{Schwerdtfeger2019} & $--$ & $--$ &   &     $11.3\pm0.2$ &     $37.3\pm 0.7$ &        $40.0\pm1$ &        $53.0\pm6$ &        $47.0\pm3$ &        $31.0\pm4$ \\
\midrule
 \multicolumn{3}{c}{\textbf{Group 15}}     &       &          N &                 P &                As &                Sb &                Bi &                Mc \\
\midrule
NR & $^4S$ & CCSD(T) &         &    $7.25$ &           $24.92$ &  $29.83 \pm 0.01$ &  $43.78 \pm 0.01$ &  $51.18 \pm 0.01$ &  $63.63 \pm 0.03$ \\
SR$_n$ & $^4S$ & CCSD(T) &     &       $7.24$ &           $24.93$ &  $29.70 \pm 0.01$ &  $42.96 \pm 0.01$ &  $48.60 \pm 0.01$ &  $60.65 \pm 0.03$ \\
SR$_d$ & $^4S$ & CCSD(T) &     &      $7.27$ &           $25.16$ &  $29.62 \pm 0.01$ &           $43.05$ &  $48.78 \pm 0.01$ &  $60.96 \pm 0.03$ \\
DC & $^4S_{3/2}$ & MRCISD &    &      $7.20$ &           $24.83$ &           $28.98$ &           $41.98$ &           $46.57$ &           $68.64$ \\
                                 & $^4S_{3/2}, M_J=1/2$ & MRCISD &   &         $7.20$ &           $24.83$ &           $29.00$ &           $42.22$ &           $51.42$ &           $97.20$ \\
                                 & $^4S_{3/2}, M_J=3/2$ & MRCISD &   &        $7.20$ &           $24.83$ &           $28.96$ &           $41.73$ &           $41.71$ &           $40.07$ \\
Rec. & $--$ & $--$ &   & $7.24 \pm 0.15$ &  $24.93 \pm 0.42$ &  $29.70 \pm 0.66$ &  $41.98 \pm 1.10$ &  $46.57 \pm 2.23$ &  $68.64 \pm 7.69$ \\
Ref.~\citenum{Schwerdtfeger2019} & $--$ & $--$ &   &     $7.4\pm0.2$ &        $25.0\pm1$ &          $30\pm1$ &          $43\pm2$ &          $48\pm4$ &       $71.0\pm20$ \\
\midrule
\multicolumn{3}{c}{\textbf{Group 16}}       &   &              O &                 S &                Se &                Te &                Po &                 Lv \\
\midrule
NR & $^3P$ & CCSD(T) &        &    $5.16$ &  $19.17 \pm 0.03$ &  $25.62 \pm 0.27$ &  $38.56 \pm 0.28$ &  $45.83 \pm 0.02$ &   $58.72 \pm 0.01$ \\
SR$_n$ & $^3P$ & CCSD(T) &    &       $5.15$ &  $19.21 \pm 0.02$ &  $25.57 \pm 0.19$ &  $38.01 \pm 0.09$ &           $44.32$ &   $57.00 \pm 0.01$ \\
SR$_d$ & $^3P$ & CCSD(T) &    &      $5.16$ &  $19.35 \pm 0.02$ &  $25.53 \pm 0.15$ &  $37.94 \pm 0.14$ &           $44.48$ &   $57.22 \pm 0.01$ \\
DC & $^3P_2$ & MRCISD &       &     $5.13$ &           $19.08$ &           $24.48$ &           $36.58$ &           $43.22$ &            $71.66$ \\
                                 & $^3P_2, M_J=0$ & MRCISD &   &         $5.41$ &           $20.21$ &           $25.87$ &           $38.00$ &           $43.83$ &            $68.41$ \\
                                 & $^3P_2, M_J=1$ & MRCISD &   &        $4.84$ &           $17.94$ &           $23.08$ &           $35.16$ &           $42.61$ &            $74.92$ \\
                                 & $^3P_2, M_J=2$ & MRCISD &   &       $5.27$ &           $19.64$ &           $25.18$ &           $37.29$ &           $43.52$ &            $70.04$ \\
Rec. & $--$ & $--$ & &  $5.15 \pm 0.04$ &  $19.21 \pm 0.32$ &  $25.57 \pm 1.08$ &  $36.58 \pm 1.37$ &  $43.22 \pm 1.28$ &  $71.66 \pm 14.45$ \\
Ref.~\citenum{Schwerdtfeger2019} & $--$ & $--$ &   &     $5.3\pm0.2$ &      $19.4\pm0.1$ &      $28.9\pm1.0$ &        $38.0\pm4$ &        $44.0\pm4$ &               $--$ \\
\midrule
\multicolumn{3}{c}{\textbf{Group 17}}    &   &              F &                Cl &                Br &                 I &                At &                 Ts \\
\midrule
NR & $^2P$ & CCSD(T) &       &      $3.60$ &           $14.54$ &  $20.57 \pm 0.22$ &  $33.20 \pm 0.96$ &           $39.50$ &   $50.26 \pm 0.12$ \\
SR$_n$ & $^2P$ & CCSD(T) &   &        $3.61$ &  $14.56 \pm 0.01$ &  $21.00 \pm 0.30$ &  $32.06 \pm 0.06$ &           $38.39$ &   $49.91 \pm 0.35$ \\
SR$_d$ & $^2P$ & CCSD(T) &   &       $3.60$ &           $14.68$ &  $20.69 \pm 0.27$ &  $32.07 \pm 0.06$ &           $38.48$ &   $50.06 \pm 0.35$ \\
DC & $^2P_{3/2}$ & MRCISD &  &       $3.57$ &           $14.36$ &           $20.06$ &           $31.28$ &           $38.92$ &            $65.15$ \\
                                 & $^2P_{3/2}, M_J=3/2$ & MRCISD &    &         $3.70$ &           $14.94$ &           $20.95$ &           $32.63$ &           $41.59$ &            $71.89$ \\
                                 & $^2P_{3/2}, M_J=1/2$ & MRCISD &    &        $3.43$ &           $13.78$ &           $19.16$ &           $29.92$ &           $36.25$ &            $58.40$ \\
                                 & $^2P_{1/2}, M_J=1/2$ & MRCISD &    &       $3.57$ &           $14.40$ &           $20.27$ &           $32.01$ &           $40.79$ &           $227.28$ \\
Rec. & $--$ & $--$ &   & $3.61 \pm 0.04$ &  $14.56 \pm 0.35$ &  $21.00 \pm 0.77$ &  $31.28 \pm 0.81$ &  $38.92 \pm 0.49$ &  $65.15 \pm 15.09$ \\
Ref.~\citenum{Schwerdtfeger2019} & $--$ & $--$ &  &   $3.74\pm0.08$ &      $14.6\pm0.1$ &        $21.0\pm1$ &      $32.9\pm1.3$ &        $42.0\pm4$ &        $76.0\pm15$ \\
\midrule
\multicolumn{3}{c}{\textbf{Group 18}}     &      He &               Ne &                Ar &                Kr &                Xe &                Rn &                Og \\
\midrule
NR & $^1S$ & CCSD(T) &  $1.38$ &           $2.57$ &           $11.01$ &           $16.78$ &  $27.23 \pm 0.01$ &  $34.08 \pm 0.01$ &  $45.26 \pm 0.02$ \\
SR & $^1S$ & CCSD(T) &  $1.38$ &           $2.57$ &           $11.02$ &           $16.77$ &           $27.01$ &           $33.12$ &           $43.74$ \\
DC & $^1S_0$ & CCSD(T) &  $1.38$ &           $2.57$ &           $11.02$ &           $16.80$ &           $27.20$ &           $35.07$ &           $58.24$ \\
Rec. & $--$ & $--$ &  $1.38$ &  $2.57 \pm 0.09$ &  $11.02 \pm 0.06$ &  $16.80 \pm 0.02$ &  $27.20 \pm 0.14$ &  $35.07 \pm 0.21$ &  $58.24 \pm 0.32$ \\
Ref.~\citenum{Schwerdtfeger2019} & $--$ & $--$ &  $1.38$ &           $2.66$ &           $11.08$ &    $16.78\pm0.02$ &     $27.32\pm0.2$ &        $35.0\pm2$ &        $58.0\pm6$ \\
\bottomrule
\end{longtable}
}

\subsection{Group 1 elements}
\label{subsec:group-1-elements}

As shown in \autoref{tab:dipole_summary}, the DC values show good agreement with the recommended values from Ref.~\citenum{Schwerdtfeger2019} for rubidium (Rb, $Z=37$) and Fr, but not for lithium (Li, $Z=3$), sodium (Na, $Z=11$), K, and Cs.

For Li, the DC result (164.31) aligns closely with experimental measurements ($164.0 \pm 3.4$~\cite{Molof1974} and $164.2 \pm 1.1$~\cite{Miffre2006a}).
The recommended value ($164.1125 \pm 0.0005$) in Ref.~\citenum{Schwerdtfeger2019} is based on the value reported in Ref.~\citenum{Puchalski2011}.
The small deviation of only 0.12\% from the recommended value may stem from the exclusion of relativistic effects such as Breit interactions and QED.

For Na, the DC result ($163.27 \pm 0.09$) is in close agreement with the experimental value ($161 \pm 7.5$) reported by Ma \textit{et al}.\cite{Ma2015}
Additionally, it slightly exceeds the recommended computational value (162.88) from Refs.~\citenum{Maroulis2006,Thakkar2005}, which were obtained using a composite scheme developed by Thakkar and Lupinetti.
The NR and SR values for Na, calculated as 163.98 and 162.99 respectively, match well with the corresponding computational values of 163.90 and 162.90 reported in Refs.~\citenum{Maroulis2006,Thakkar2005}.
The difference between the DC value in this work and the final recommended value from Refs.~\citenum{Maroulis2006,Thakkar2005} can be attributed to the method used to account for the SOC contribution to $\alpha$.
In Refs.~\citenum{Maroulis2006,Thakkar2005}, a value of $-0.02$ a.u. from Ref.~\citenum{Lim1999}, calculated at the DHF level, was added to the SR value.
In contrast, the SOC contribution in this work was evaluated at the CCSD(T) level as $+0.28$ a.u.
It is noteworthy that the SOC contributions at the DHF and CCSD levels in this study are $-0.04$ and $-0.03$ a.u., respectively.
Additionally, this study confirms that the NR value is consistent with the recent non-relativistic CCSD(T) value (163.9) reported by Smiałkowski and Tomza at the CCSD(T)/aug-cc-pwCV5Z level,\cite{Smialkowski2021} further supporting the reliability of the method employed.

For K, the lower bound of the DC value ($290.64 \pm 0.24$) is marginally higher than the upper bound of the recommended value ($289.7 \pm 0.3$),\cite{Schwerdtfeger2019} which is based on experimental data.\cite{Gregoire2015,Gregoire2016}
However, the DC value obtained using CCSD(T) from Eq.~\eqref{eq:ff_d_1} is $289.40 \pm 0.20$ (as shown in Table S9), in good agreement with the recommended value from Ref.~\citenum{Schwerdtfeger2019}.
The slight discrepancy may arise from the invalidity of a negative $\gamma$ parameter obtained from Eq.~\eqref{eq:ff_d_2}, particularly when $\tau > 0.1\%$ is used in Eq.~\eqref{eq:tau}.

For Cs, the DC value ($398.24 \pm 0.34$) is slightly lower than the value of 399.0 reported by Borschevsky \textit{et al.},\cite{Borschevsky2013} which was calculated using the CCSD(T) method with the F{\ae}gri basis set.\cite{FaegriJr2001}
This minor discrepancy could be due to differences in the basis set and fitting procedures used in this study.
Further details can be found in the Supporting Information.

\begin{figure}[h]
\centering
\includegraphics[scale=1]{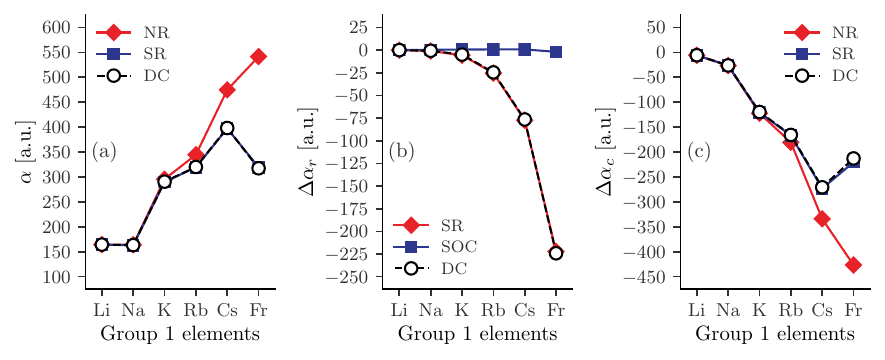}
\caption{
Polarizabilities (in a.u.) of Group 1 elements.
(a) Comparison of non-relativistic (NR), scalar-relativistic (SR), and Dirac-Coulomb (DC) dipole polarizabilities.
(b) Illustration of the influence of relativistic effects, including SR, spin-orbit coupling (SOC), and DC, on dipole polarizabilities.
(c) Examination of the impact of electron correlation on dipole polarizabilities in the presence of various relativistic effects.
}
\label{fig:group_1}
\end{figure}

\autoref{fig:group_1}(a) demonstrates that non-relativistic polarizabilities increase monotonically with atomic numbers, starting from Na.
In contrast, the relativistic data increase with increasing atomic numbers from Na to Cs, but decrease from Cs to Fr, with the value for Fr being close to that of Rb.
This trend confirms the irregular progression of values with atomic numbers for Group 1 elements, driven by the relativistic contraction of the valence $s$ shell.\cite{Lim1999,Lim2005}
The polarizability of Na is slightly lower than that of Li due to reduced screening of the nuclear charge by the $p$ electrons in Na, resulting in more compact valence shells and lower polarizabilities.

\autoref{fig:group_1}(b) indicates that the scalar-relativistic effect is more pronounced than the SOC effect, leading to a roughly $Z^2$ increase in relativistic effects on dipole polarizabilities due to the scalar-relativistic contribution.
The SOC contribution to dipole polarizabilities for Group 1 atoms is small; generally, the SOC effect slightly increases the polarizabilities of Group 1 elements, except for Fr, where the SOC effect reduces dipole polarizability by nearly 2.
This observation aligns with previous findings.\cite{Lim1999,Lim2005}

As depicted in \autoref{fig:group_1}(c), at the non-relativistic level, the electron correlation contribution increases monotonically in absolute value with increasing atomic numbers.
In contrast, relativistic calculations exhibit the same trend up to Cs, although the contributions from Rb and Cs are slightly lower than their non-relativistic counterparts.
An unusual trend is observed for Fr, where the electron correlation contributions in relativistic calculations are significantly lower than those in the non-relativistic case.

 \subsection{Group 2 elements}
\label{subsec:group-2-elements}

As shown in \autoref{tab:dipole_summary}, the DC values are in good agreement with the recommended values, except for beryllium (Be, $Z=4$) and Sr, where the DC values slightly exceed the recommended upper bounds.
However, the DC result (37.78) for Be using CCSD(T) aligns well with recent theoretical values ($37.73 \pm 0.04$~\cite{Jiang2015a, Cheng2013} and 37.787~\cite{Wu2023}) from the RCICP method, as well as the CI+MBPT2 result ($37.76 \pm 0.22$) based on experimental data.\cite{Porsev2006}
In addition, the recommended value ($197.2 \pm 0.2$) for Sr in Ref.~\citenum{Schwerdtfeger2019} is based on computational results from Refs.~\citenum{Porsev2006, Derevianko2010}, where a hybrid relativistic CI and MBPT method is used.
This implies that the discrepancy may arise from higher-order correlation effects included in this work.

\begin{figure}[h]
    \centering
    \includegraphics[scale=1]{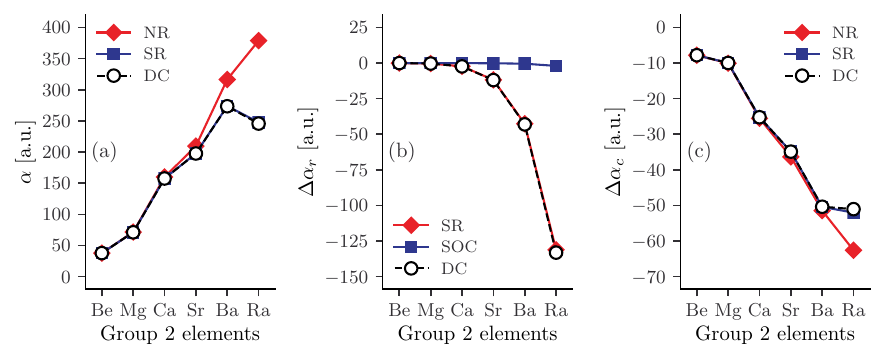}
    \caption{
        Same as \autoref{fig:group_1}, but for Group 2 elements.
    }
    \label{fig:group_2}
\end{figure}

\autoref{fig:group_2}(a) displays the relationship between dipole polarizabilities and atomic numbers for Group 2 elements.
It reveals that non-relativistic polarizabilities increase monotonically with atomic numbers.
However, the trend for relativistic polarizabilities resembles the non-relativistic case for lighter atoms but is lower for heavier elements, particularly for Sr and barium (Ba, $Z=56$), where deviations exceed 11 and 42, respectively.
Moreover, a considerable decrease in relativistic polarizabilities is observed from Ba to radium (Ra, $Z=88$).

\autoref{fig:group_2}(b) highlights that the scalar-relativistic effect is more significant than the SOC effect, leading to an approximate $Z^2$ increase in the relativistic effects on the scalar-relativistic contribution to dipole polarizabilities, paralleling the trend observed in Group 1 elements.
The SOC contribution to dipole polarizabilities for Group 2 atoms is small.

\autoref{fig:group_2}(c) indicates that the electron correlation contribution in all relativistic calculations grows monotonically in absolute value with increasing atomic numbers, except for Ra atom where the electron-correlation contributions of relativistic calculations are nearly 10 less than in the non-relativistic simulations.

 \subsection{Group 13 elements}
\label{subsec:group-13-elements}

As shown in \autoref{tab:dipole_summary}, both DC-CC and DC-CI values in this work align well with the recommended theoretical values presented in Ref.~\citenum{Schwerdtfeger2019}, except for boron (B, $Z=5$) and Nh.
It should be noted that the DC-CC data for B is not available in the present work due to convergence issues encountered with DHF using a DC Hamiltonian.
Therefore, the most accurate results are DC-CC results for all Group 13 elements except for B, for which the SR value is used as the most accurate result.
For Al, the best DC-CC result is obtained by including 11 outer-shell electrons and 223 virtual shells for electron correlation instead of all electrons and virtual shells, as shown in Table S3.
Furthermore, all component-resolved values for Group 13 atoms are consistent with the results computed in Ref.~\citenum{Fleig2005}.
While an earlier study has shown that virtual cutoff values of 10 are sufficient for DC calculations,\cite{Fleig2005} I opted for a tight cutoff of 25 in this work.
This choice stems from the confidence in earlier findings and the anticipation that a higher cutoff might ensure augmented accuracy.

For Nh, the computed DC-CC dipole polarizability of $32.89 \pm 0.18$ exceeds the recommended value of $29.2 \pm 2$.
Prior studies using FS-CCSD (29.85) and SD+CI (28.8) reported lower values.
Differences arise from the larger basis set and fewer correlated electrons in this work.
Validation tests, including two-component DC-CC calculations and a larger basis set, confirm the convergence of the basis set and number of correlated electrons.
The computed DC-CI value of $28.21$ agrees with prior SD+CI results.
This suggests an updated recommended value for Nh based on the CCSD(T) result is warranted.
Further details can be found in the Supporting Information.

\begin{figure}[h]
\centering
\includegraphics[scale=1]{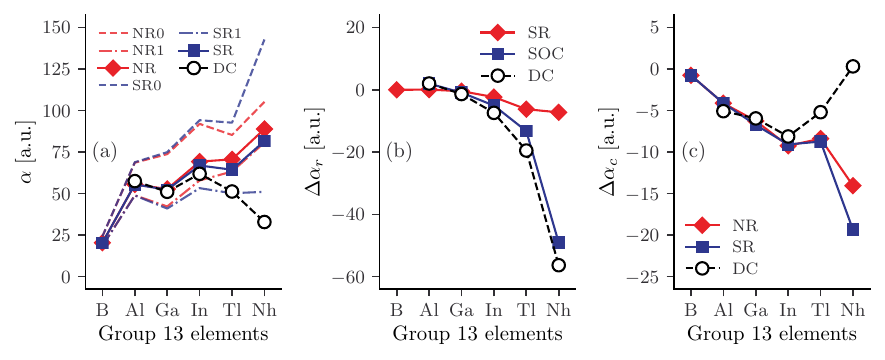}
\caption{
Polarizabilities (in a.u.) of Group 13 elements.
(a) Comparison of non-relativistic (NR), scalar-relativistic (SR), and Dirac-Coulomb (DC) dipole polarizabilities.
The results of states $M_L=0$ and $M_L=\pm1$ are indicated as NR0 (SR0) and NR1 (SR1) for non-relativistic (scalar-relativistic) calculations, respectively.
(b) Illustration of the influence of relativistic effects, including SR, SOC, and DC, on dipole polarizabilities.
(c) Examination of the impact of electron correlation on dipole polarizabilities in the presence of various relativistic effects.
}
\label{fig:group_13}
\end{figure}

\autoref{fig:group_13}(a) displays the relationship between dipole polarizabilities and atomic numbers for Group 13 elements.
Also, the $M_L$ component polarizabilities at the spin-free-relativistic level are presented.
In general, polarizabilities increase with atomic numbers at the non-relativistic level, with the exception of Ga, which has a smaller screening of the nuclear charge by $d$ electrons, resulting in more compact valence shells and a reduced polarizability as reported in the previous study.\cite{Fleig2005}
From the component-resolved results in \autoref{fig:group_13}(a), the decrease in polarizability is mainly caused by the $M_L=\pm1$ state.
This effect is also observed in the comparison between In and Tl, but it is smaller than in the Ga case at the non-relativistic level.
The difference in $M_L=0$ states of In and Tl eliminates the increase in the corresponding $M_L=\pm1$ components, resulting in a polarizability value for Tl that is close to that of In in non-relativistic calculations.
However, at the scalar-relativistic level, this change cannot be attributed solely to $f$ electrons screening due to relativity.
In the Dirac picture, the SOC effect significantly reduces the polarizabilities of Tl and Nh.
As a result, the polarizability of Tl is close to that of Ga, and the result for Nh is lower than Al.
This finding is consistent with previous work in Ref.~\citenum{Fleig2005}.

Both relativistic effects and electron correlation have a reducing effect on the polarizabilities of Group 13 elements, as seen in Figures~\ref{fig:group_13}(b) and (c).
The DC relativistic effects ($\Delta \alpha_r^\text{DC}$) roughly increase with atomic numbers to the power of 2, as well as through the SOC effect, which is the dominant factor compared to scalar-relativistic effects ($\Delta \alpha_r^\text{SR}$).
An exception is Al, whose DC-CC calculation is not accurate.
The smaller difference between $\bar{\alpha}_{J=3/2}$ and $\alpha_{J=1/2}$ for B and Al in \autoref{tab:dipole_group_13} also suggests that SOC contribution can be disregarded entirely, which is consistent with previous research presented in Ref.~\citenum{Fleig2005}.

In \autoref{fig:group_13}(c), the electron correlation effects generally increase with increasing atomic numbers at the non-relativistic and scalar-relativistic levels, with the exception of Tl, whose $\Delta \alpha_c^\text{NR}$ and $\Delta \alpha_c^\text{SR}$ are slightly larger than those of In, respectively.
In the Dirac picture, electron correlation plays a critical role for Ga and In, whereas its effects decrease with increasing atomic numbers starting with In due to the rising influence of the SOC effect.

 \subsection{Group 14 elements}
\label{subsec:group-14}

As shown in \autoref{tab:dipole_summary}, all DC results are in good agreement with the recommended values in Ref.~\citenum{Schwerdtfeger2019} except for carbon (C, $Z=6$).
Additionally, the DC value (11.73) of C is slightly larger than the upper bound of the result ($11.26 \pm 0.20$) reported in Ref.~\citenum{Thierfelder2008}.
The difference could be attributed to the different basis sets and the number of data points computed.
The basis set ($19s11p6d4f2g$) is used in this work compared to the ($13s7p4d3f2g$) used in Ref.~\citenum{Thierfelder2008}.
In addition, energy with an external electric field of $+0.0005$ a.u. is also added in this work compared to Ref.~\citenum{Thierfelder2008}.
Moreover, the difference between SR and DC values, as shown in \autoref{tab:dipole_summary}, could be attributable to the small SOC effect for carbon.
This suggests that $LS$ coupling may provide a better description than $jj$ coupling.
The calculated SR polarizability of C (11.61) is close to the DK value (11.70) in Ref.~\citenum{Thierfelder2008} and the recent scalar-relativistic CCSD(T) result (11.524).\cite{CanalNeto2021}
Due to the small relativistic effects, numerous NR values ($11.67 \pm 0.07$~\cite{Das1998}, 11.63~\cite{A.Manz2019}, $11.64 \pm 0.15$~\cite{Wang2021}, 11.683~\cite{Ehn2021}, 11.324~\cite{Ehn2021}) computed by the CCSD(T) method have been reported in the literature.
The computed NR value (11.66) in this work is in good agreement with these values, with the largest error being less than 2\%.

\begin{figure}[htbp]
    \centering
    \includegraphics[scale=1.0]{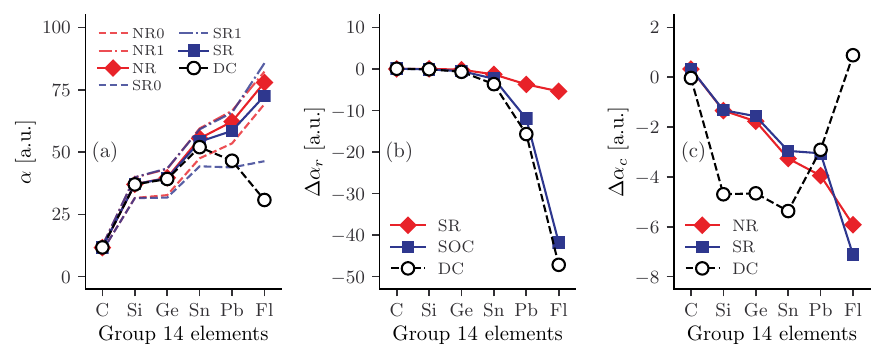}
    \caption{
        Same as \autoref{fig:group_13}, but for Group 14 elements.
    }
    \label{fig:group_14}
\end{figure}

The relationship between dipole polarizabilities and atomic numbers for Group 14 elements is depicted in \autoref{fig:group_14}(a).
In general, all dipole polarizabilities increase with atomic numbers except for atoms from tin (Sn, $Z=50$) to flerovium (Fl, $Z=114$) due to the SOC effect.
\autoref{fig:group_14}(b) shows that $\Delta \alpha_r^\text{SOC}$ and $\Delta \alpha_r^\text{DC}$ increase approximately as $Z^2$ with atomic numbers, while $\Delta \alpha_r^\text{SR}$ increases linearly as $Z$ with respect to atomic numbers.
Moreover, the main difference between $\alpha^\text{NR}$ and $\alpha^\text{SR}$ is from the $M_L=0$ component, and the difference increases with atomic numbers, especially lead (Pb, $Z=82$) and Fl.
For lighter atoms C, silicon (Si, $Z=14$), and germanium (Ge, $Z=32$), all results are very close, suggesting that relativistic effects are negligible.

As shown in \autoref{fig:group_14}(c), electron correlation reduces the dipole polarizabilities except for the DC value of Fl, which agrees with the previous findings in Ref.~\citenum{Thierfelder2008}.
The large deviation between $\Delta \alpha_c^\text{SR}$ and $\Delta \alpha_c^\text{DC}$ could be attributed to the fact that the full-relativistic DC DHF wave function is not a proper reference wave function, leading to considerable correlation energy for Si, Ge, and Sn.

 \subsection{Group 15 elements}
\label{subsec:group_15}

As shown in \autoref{tab:dipole_summary}, all DC, SR, and NR polarizabilities of the ground state for Group 15 elements are in excellent agreement with the recommended values reported in Ref.~\citenum{Schwerdtfeger2019}, except for the DC value of As (28.98), which is slightly smaller than the lower bound of the corresponding recommended value ($30 \pm 1$).
This implies that the relativistic effects are small for all Group 15 elements except for Mc, where a large uncertainty is used.
Therefore, in this work, the most accurate calculations for nitrogen (N, $Z=7$), phosphorus (P, $Z=15$), and arsenic (As, $Z=33$) are SR CCSD(T), while DC MRCISD is used for antimony (Sb, $Z=51$), bismuth (Bi, $Z=83$), and Mc due to the increasing SOC contribution to polarizabilities.

There are no DC values available for N, P, and As yet, based on Ref.~\citenum{Schwerdtfeger2019,Schwerdtfeger2023}.
For Sb, only one DC value ($45 \pm 11$) is reported by Doolen with relativistic local density approximation (LDA) in linear response theory.\cite{Doolen1987}
Additionally, this method is also used to compute the DC value ($50 \pm 12$) of Bi.
The most accurate DC values for Bi and Mc in the literature are reported by Dzuba \textit{et al.} using SD+CI methods.
It should be noted that the DC values for Bi and Mc in this work are 46.57 and 68.64, respectively, which are in excellent agreement with the SD+CI values (44.62 and 70.5).\cite{Dzuba2016}

\begin{figure}[h]
    \centering
    \includegraphics[scale=1.0]{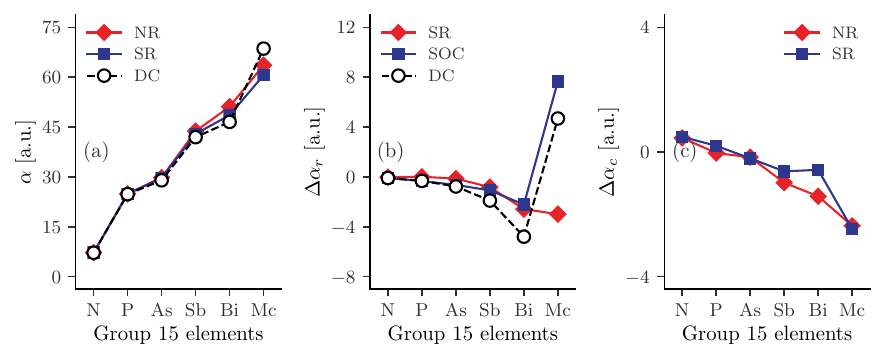}
    \caption{
        Same as \autoref{fig:group_1}, but for Group 15 elements.
    }
    \label{fig:group_15}
\end{figure}

The relationship between dipole polarizabilities and atomic numbers for Group 15 elements is depicted in \autoref{fig:group_15}(a).
It can be seen that the dipole polarizability generally increases with atomic number.
Additionally, the SOC effects are as crucial as scalar-relativistic effects starting from Sb, as shown in \autoref{fig:group_15}(b).
Both $\Delta \alpha_c^\text{NR}$ and $\Delta \alpha_c^\text{SR}$ increase with atomic numbers and reduce polarizabilities, starting from P.
Additionally, relativistic effects generally reduce polarizabilities, with the exception of Mc, where the contribution of the SOC effect significantly improves the polarizability by nearly 8.0.
\autoref{fig:group_15}(c) suggests that the electron correlation effects are small in both non-relativistic and scalar-relativistic calculations.

 \subsection{Group 16 elements}
\label{subsec:group-16-elements}

As shown in \autoref{tab:dipole_summary}, the DC-CI results are in agreement with the recommended values presented in Ref.~\citenum{Schwerdtfeger2019}, except for sulfur (S, $Z=16$) and selenium (Se, $Z=34$).
Considering the small relativistic effects for light elements, the most accurate calculations for oxygen (O, $Z=8$), S, and Se in this work are SR CCSD(T).
In contrast, DC MRCISD is used for tellurium (Te, $Z=52$), polonium (Po, $Z=84$), and Lv due to the increasing SOC contribution to polarizabilities.

\begin{figure}[h]
    \centering
    \includegraphics[scale=1.0]{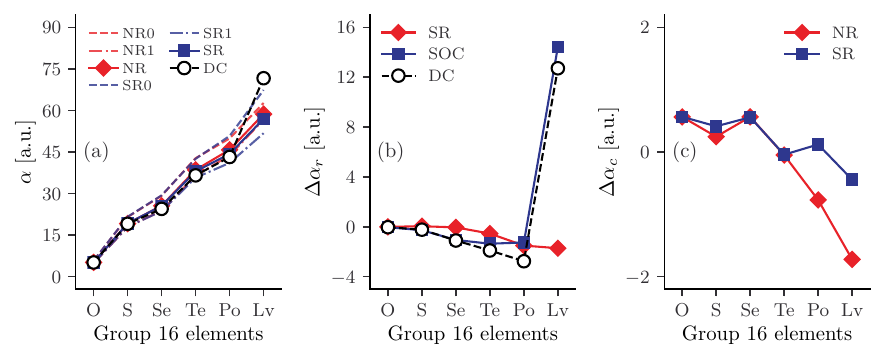}
    \caption{
        Same as \autoref{fig:group_13}, but for Group 16 elements.
    }
    \label{fig:group_16}
\end{figure}

The relationship between dipole polarizabilities and atomic numbers for Group 16 elements is presented in \autoref{fig:group_16}(a).
This figure also highlights the $M_L=0$ and $M_L=\pm1$ polarizabilities, denoted as NR0 (SR0) and NR1 (SR1) for non-relativistic (scalar-relativistic) calculations, respectively.
It is observed that the polarizabilities slightly increase with atomic numbers, and a significant discrepancy between scalar-relativistic and DC polarizabilities occurs in Lv due to the SOC effect.
From \autoref{fig:group_16}(b), relativistic effects slightly reduce polarizabilities, but in the case of Lv, the SOC effects notably increase the polarizability.
Lastly, \autoref{fig:group_16}(c) shows that electron correlation marginally enhances polarizabilities in atoms up to Te but reduces them in heavier atoms beyond Te.

 \subsection{Group 17 elements}
\label{subsec:group-17-elements}

As shown in \autoref{tab:dipole_summary}, the DC values align well with the recommended values, except for the light atoms, F and chlorine (Cl, $Z=17$).
Given the small relativistic effects in light elements, the most accurate calculations for F, Cl, and Br in this work are SR CCSD(T).
In contrast, DC MRCISD is used for I, At, and Ts due to the increasing SOC contribution to polarizabilities.
As expected, the polarizabilities of states $ J=\frac{3}{2}, M_J=\frac{1}{2}$ are found to be smaller than those of the state $J=\frac{3}{2}, M_J=\frac{3}{2}$.
Moreover, the values for $J=\frac{3}{2}, M_J=\frac{3}{2}$ are also slightly larger than those for $J=\frac{1}{2}, M_J=\frac{1}{2}$ across all elements except for Ts, which displays an anomalously high polarizability, nearly four times larger than that of the $\alpha_{3/2,1/2}$ state.
This deviation can be attributed to the heavily relativistic contraction of the $\phi\left(\frac{1}{2}, \frac{1}{2}\right)$ spinors, which are less polarizable and more occupied in the $J=\frac{3}{2}$ states as reported in Ref.~\citenum{Fleig2002}.
The small values of $\alpha_{J=1/2}-\bar{\alpha}_{J=3/2}$ and $\bar{\alpha}_{M_L}-\bar{\alpha}_{3/2}$ suggest that the SOC effect can be neglected in the calculation of dipole polarizabilities for Group 17 atoms, except for Ts, where the SOC effects are particularly significant.

\begin{figure}[h]
    \centering
    \includegraphics[scale=1.0]{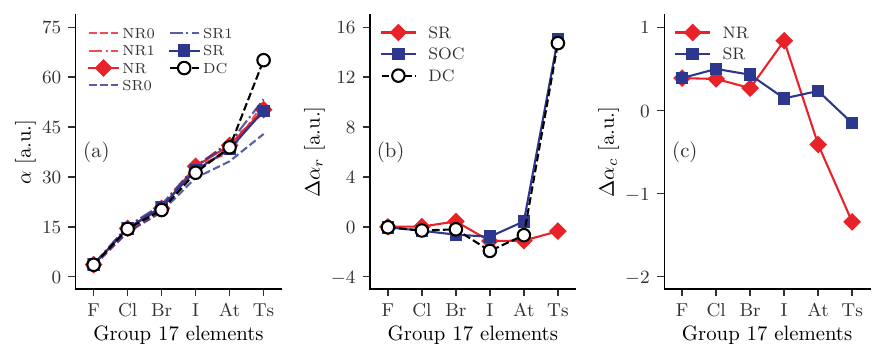}
    \caption{
        Same as \autoref{fig:group_13}, but for Group 17 elements.
    }
    \label{fig:group_17}
\end{figure}

The relationship between dipole polarizabilities and atomic numbers for Group 17 elements is presented in \autoref{fig:group_17}(a).
This figure also highlights the $M_L=0$ and $M_L=\pm1$ polarizabilities, denoted as NR0 (SR0) and NR1 (SR1) for non-relativistic (scalar-relativistic) calculations, respectively.
All results indicate that dipole polarizabilities increase with atomic numbers.
\autoref{fig:group_17}(b) provides the group trend of relativistic effects on the static polarizabilities of Group 17 atoms.
It is observed that the SOC effect significantly improves polarizabilities for the element Ts, indicating that the SOC effects are particularly significant for Ts.
The electron correlation contribution is presented in \autoref{fig:group_17}(c), indicating that electron correlation is relatively small, with values less than 2 for non-relativistic calculations and less than 1 for scalar-relativistic calculations.
For light atoms, electron correlation slightly increases polarizabilities at both non-relativistic and scalar-relativistic levels, whereas for heavy atoms, it reduces polarizabilities.

 \subsection{Group 18 elements}
\label{subsec:group-18-elements}

As shown in \autoref{tab:dipole_summary}, the DC results for the heavy atoms, including krypton (Kr, $Z=36$), xenon (Xe, $Z=54$), radon (Rn, $Z=86$), and Og, show excellent agreement with the reference values in Ref.~\citenum{Schwerdtfeger2019}.
For the lighter atoms, like Ne, and argon (Ar, $Z=18$), the polarizabilities calculated in this study are slightly lower than the recommended values as shown in \autoref{tab:dipole_summary}.
Moreover, the NR, SR, and DC values are 2.57 for Ne in this work, which is lower than experimental~\cite{Huot1991,Dalgarno1997b,Gaiser2010,Gaiser2018} and theoretical values,\cite{Soldán2001a,Maroulis2006,Lesiuk2020,Wang2021,Hellmann2022,Mori2023} i.e., 2.66.
Additionally, the SR and DC values are 11.02 for Ar in this work, which is lower than experimental~\cite{Newell1965,Orcutt1967,Dalgarno1997b,Maroulis2006,Gaiser2018} and theoretical values,\cite{Maroulis2006,Hohm2012,Lupinetti2005,Lesiuk2023} i.e., 11.08.

One possible reason for the discrepancy is the use of basis sets, which might not be sufficient to accurately describe the polarization in these light atoms.
Another possible reason could be attributed to the use of the Gaussian model as the nucleus model, while a point nucleus model is usually employed in nonrelativistic programs.

\begin{figure}[h]
    \centering
    \includegraphics[scale=1.0]{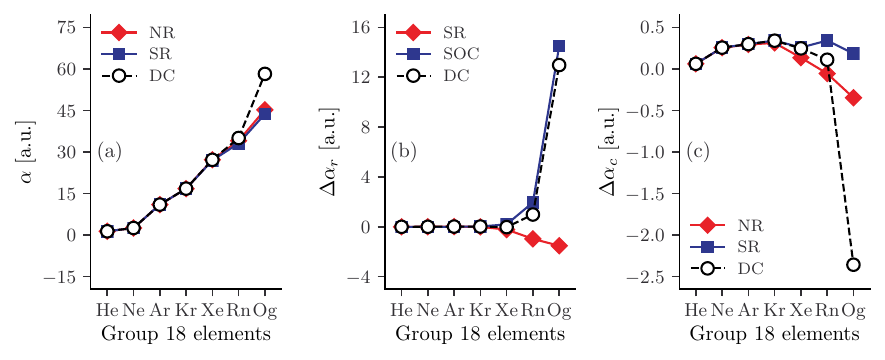}
    \caption{
        Same as \autoref{fig:group_1}, but for Group 18 elements.
    }
    \label{fig:group_18}
\end{figure}

The relationship between dipole polarizabilities and atomic numbers for Group 18 elements is depicted in \autoref{fig:group_18}(a), indicating that both non-relativistic and relativistic values increase with atomic numbers.
The contributions of relativistic effects are depicted in \autoref{fig:group_18}(b).
The relativistic effects are small in light atoms like He, Ne, Ar, and Kr.
As expected, the contribution of the SOC effect increases with atomic numbers, starting from Xe (0.20), increasing at Rn (1.95), and reaching a maximum for the Og atom (14.50).
This finding suggests that the contribution of the SOC effect is indispensable for polarizability calculations of heavy atoms, especially for Og.
\autoref{fig:group_18}(c) displays the electron correlation effects, which are generally small ($<0.5$) for Group 18 elements.
However, Og is an exception with $\Delta \alpha_c^\text{DC} = -2.36$.

    \section{Summary}
    \label{sec:summary}
    In this study, I obtained the static dipole polarizabilities of main-group elements using the finite-field method combined with relativistic CCSD(T) and MRCI calculations.
I also provide a systematic analysis of the impact of various relativistic effects on atomic dipole polarizabilities, including scalar-relativistic effects, spin-orbit coupling (SOC), and the full Dirac-Coulomb relativistic effect.
For heavy atoms ($Z>40$), my theoretical values align well with the recommended values.
Moreover, the analysis suggests that scalar-relativistic effects are predominant for elements in Groups 1--2, whereas SOC effects are typically negligible in these calculations.
For elements in Groups 13--18, scalar-relativistic effects contribute minimally to dipole polarizability.
However, the influence of SOC becomes significant starting from the fourth row of the Periodic Table for elements in Groups 13--14 and from the fifth row for elements in Groups 15--18.
Additionally, the influence of electron correlation in the presence of various relativistic effects is investigated.
Generally, it is found to be crucial for elements in Groups 1--2 and significant for those in Groups 13--14.
However, for elements in Groups 15--18, its contribution is small.
This study, employing computationally accurate methods with extensively large basis sets, yields comprehensive and consistent dipole polarizabilities for main-group elements, particularly for heavy atoms where only empirical values exist.

    \begin{acknowledgement}
            Y.C. acknowledges the Foundation of Scientific Research - Flanders (FWO, file number G0A9717N) and the Research Board of Ghent University (BOF) for their financial support.
    The resources and services used in this work were provided by the VSC (Flemish Supercomputer Center), funded by the Research Foundation - Flanders (FWO) and the Flemish Government.
    The author thanks Prof. Dr. Toon Verstraelen, and Dr. Thiago Carvalho Corso for helpful comments on the manuscript.
     \end{acknowledgement}

    \begin{suppinfo}
        The Supplementary Material includes a PDF document and a file in ``CSV'' format.
The PDF document contains the following information:
\begin{enumerate}
    \item Additional detailed atomic dipole polarizabilities obtained by fitting Eq.~\eqref{eq:ff_d_2}, including those for DHF, MP2, and CCSD, as well as component-resolved results for open-shell atoms.
    \item The atomic dipole polarizabilities obtained by fitting Eq.~\eqref{eq:ff_d_1} are also included.
        A summary table of the most accurate results for all main groups is provided.
    \item The reference atomic dipole polarizabilities from the literature, reported in Refs.~\citenum{Schwerdtfeger2019,Schwerdtfeger2023}, are sorted by publication year for each group.
        A detailed comparison between the computational values in this work and the data from the literature is provided for each atom.
    \item A detailed uncertainty estimation is provided for each atom.
\end{enumerate}
The CSV document contains the specified external electric field used for each $\alpha$ listed in the PDF document.
     \end{suppinfo}

    \providecommand{\latin}[1]{#1}
\makeatletter
\providecommand{\doi}
  {\begingroup\let\do\@makeother\dospecials
  \catcode`\{=1 \catcode`\}=2 \doi@aux}
\providecommand{\doi@aux}[1]{\endgroup\texttt{#1}}
\makeatother
\providecommand*\mcitethebibliography{\thebibliography}
\csname @ifundefined\endcsname{endmcitethebibliography}
  {\let\endmcitethebibliography\endthebibliography}{}


\begin{mcitethebibliography}{90}
\providecommand*\natexlab[1]{#1}
\providecommand*\mciteSetBstSublistMode[1]{}
\providecommand*\mciteSetBstMaxWidthForm[2]{}
\providecommand*\mciteBstWouldAddEndPuncttrue
  {\def\EndOfBibitem{\unskip.}}
\providecommand*\mciteBstWouldAddEndPunctfalse
  {\let\EndOfBibitem\relax}
\providecommand*\mciteSetBstMidEndSepPunct[3]{}
\providecommand*\mciteSetBstSublistLabelBeginEnd[3]{}
\providecommand*\EndOfBibitem{}
\mciteSetBstSublistMode{f}
\mciteSetBstMaxWidthForm{subitem}{(\alph{mcitesubitemcount})}
\mciteSetBstSublistLabelBeginEnd
  {\mcitemaxwidthsubitemform\space}
  {\relax}
  {\relax}

\bibitem[Schwerdtfeger and Nagle(2019)Schwerdtfeger, and
  Nagle]{Schwerdtfeger2019}
Schwerdtfeger,~P.; Nagle,~J.~K. 2018 {{Table}} of Static Dipole
  Polarizabilities of the Neutral Elements in the Periodic Table. \emph{Mol.
  Phys.} \textbf{2019}, \emph{117}, 1200--1225\relax
\mciteBstWouldAddEndPuncttrue
\mciteSetBstMidEndSepPunct{\mcitedefaultmidpunct}
{\mcitedefaultendpunct}{\mcitedefaultseppunct}\relax
\EndOfBibitem
\bibitem[Bast \latin{et~al.}(2008)Bast, Heßelmann, Sałek, Helgaker, and
  Saue]{Bast2008a}
Bast,~R.; Heßelmann,~A.; Sałek,~P.; Helgaker,~T.; Saue,~T. Static and
  {Frequency}-{Dependent} {Dipole}–{Dipole} {Polarizabilities} of {All}
  {Closed}-{Shell} {Atoms} up to {Radium}: {A} {Four}-{Component}
  {Relativistic} {DFT} {Study}. \emph{ChemPhysChem} \textbf{2008}, \emph{9},
  445--453, \_eprint:
  https://onlinelibrary.wiley.com/doi/pdf/10.1002/cphc.200700504\relax
\mciteBstWouldAddEndPuncttrue
\mciteSetBstMidEndSepPunct{\mcitedefaultmidpunct}
{\mcitedefaultendpunct}{\mcitedefaultseppunct}\relax
\EndOfBibitem
\bibitem[Tang \latin{et~al.}(2018)Tang, Yu, and Dong]{Tang2018}
Tang,~Z.-M.; Yu,~Y.-M.; Dong,~C.-Z. Determination of Static Dipole
  Polarizabilities of {{Yb}} Atom*. \emph{Chinese Phys. B} \textbf{2018},
  \emph{27}, 063101\relax
\mciteBstWouldAddEndPuncttrue
\mciteSetBstMidEndSepPunct{\mcitedefaultmidpunct}
{\mcitedefaultendpunct}{\mcitedefaultseppunct}\relax
\EndOfBibitem
\bibitem[Guo \latin{et~al.}(2021)Guo, Yu, Liu, Suo, and Sahoo]{Guo2021}
Guo,~X.~T.; Yu,~Y.~M.; Liu,~Y.; Suo,~B.~B.; Sahoo,~B.~K. Electric dipole and
  quadrupole properties of the {Cd} atom for atomic-clock applications.
  \emph{Phys. Rev. A} \textbf{2021}, \emph{103}, 013109\relax
\mciteBstWouldAddEndPuncttrue
\mciteSetBstMidEndSepPunct{\mcitedefaultmidpunct}
{\mcitedefaultendpunct}{\mcitedefaultseppunct}\relax
\EndOfBibitem
\bibitem[Ludlow \latin{et~al.}(2008)Ludlow, Zelevinsky, Campbell, Blatt, Boyd,
  de~Miranda, Martin, Thomsen, Foreman, Ye, Fortier, Stalnaker, Diddams,
  Le~Coq, Barber, Poli, Lemke, Beck, and Oates]{Ludlow2008}
Ludlow,~A.~D.; Zelevinsky,~T.; Campbell,~G.~K.; Blatt,~S.; Boyd,~M.~M.;
  de~Miranda,~M. H.~G.; Martin,~M.~J.; Thomsen,~J.~W.; Foreman,~S.~M.; Ye,~J.;
  Fortier,~T.~M.; Stalnaker,~J.~E.; Diddams,~S.~A.; Le~Coq,~Y.; Barber,~Z.~W.;
  Poli,~N.; Lemke,~N.~D.; Beck,~K.~M.; Oates,~C.~W. Sr {Lattice} {Clock} at 1
  × 10–16 {Fractional} {Uncertainty} by {Remote} {Optical} {Evaluation} with
  a {Ca} {Clock}. \emph{Science} \textbf{2008}, \emph{319}, 1805--1808,
  Publisher: American Association for the Advancement of Science\relax
\mciteBstWouldAddEndPuncttrue
\mciteSetBstMidEndSepPunct{\mcitedefaultmidpunct}
{\mcitedefaultendpunct}{\mcitedefaultseppunct}\relax
\EndOfBibitem
\bibitem[Lemke \latin{et~al.}(2009)Lemke, Ludlow, Barber, Fortier, Diddams,
  Jiang, Jefferts, Heavner, Parker, and Oates]{Lemke2009}
Lemke,~N.~D.; Ludlow,~A.~D.; Barber,~Z.~W.; Fortier,~T.~M.; Diddams,~S.~A.;
  Jiang,~Y.; Jefferts,~S.~R.; Heavner,~T.~P.; Parker,~T.~E.; Oates,~C.~W.
  Spin-\$1/2\$ {Optical} {Lattice} {Clock}. \emph{Physical Review Letters}
  \textbf{2009}, \emph{103}, 063001, Publisher: American Physical Society\relax
\mciteBstWouldAddEndPuncttrue
\mciteSetBstMidEndSepPunct{\mcitedefaultmidpunct}
{\mcitedefaultendpunct}{\mcitedefaultseppunct}\relax
\EndOfBibitem
\bibitem[Ludlow \latin{et~al.}(2015)Ludlow, Boyd, Ye, Peik, and
  Schmidt]{Ludlow2015}
Ludlow,~A.~D.; Boyd,~M.~M.; Ye,~J.; Peik,~E.; Schmidt,~P. Optical atomic
  clocks. \emph{Reviews of Modern Physics} \textbf{2015}, \emph{87}, 637--701,
  Publisher: American Physical Society\relax
\mciteBstWouldAddEndPuncttrue
\mciteSetBstMidEndSepPunct{\mcitedefaultmidpunct}
{\mcitedefaultendpunct}{\mcitedefaultseppunct}\relax
\EndOfBibitem
\bibitem[Schmidt \latin{et~al.}(2007)Schmidt, Gavioso, May, and
  Moldover]{Schmidt2007}
Schmidt,~J.~W.; Gavioso,~R.~M.; May,~E.~F.; Moldover,~M.~R. Polarizability of
  {{Helium}} and {{Gas Metrology}}. \emph{Phys. Rev. Lett.} \textbf{2007},
  \emph{98}, 254504\relax
\mciteBstWouldAddEndPuncttrue
\mciteSetBstMidEndSepPunct{\mcitedefaultmidpunct}
{\mcitedefaultendpunct}{\mcitedefaultseppunct}\relax
\EndOfBibitem
\bibitem[Gaiser and Fellmuth(2010)Gaiser, and Fellmuth]{Gaiser2010}
Gaiser,~C.; Fellmuth,~B. Experimental Benchmark Value for the Molar
  Polarizability of Neon. \emph{Europhys. Lett.} \textbf{2010}, \emph{90},
  63002\relax
\mciteBstWouldAddEndPuncttrue
\mciteSetBstMidEndSepPunct{\mcitedefaultmidpunct}
{\mcitedefaultendpunct}{\mcitedefaultseppunct}\relax
\EndOfBibitem
\bibitem[Borschevsky \latin{et~al.}(2013)Borschevsky, Pershina, Eliav, and
  Kaldor]{Borschevsky2013}
Borschevsky,~A.; Pershina,~V.; Eliav,~E.; Kaldor,~U. Ab Initio Studies of
  Atomic Properties and Experimental Behavior of Element 119 and Its Lighter
  Homologs. \emph{J. Chem. Phys.} \textbf{2013}, \emph{138}, 124302\relax
\mciteBstWouldAddEndPuncttrue
\mciteSetBstMidEndSepPunct{\mcitedefaultmidpunct}
{\mcitedefaultendpunct}{\mcitedefaultseppunct}\relax
\EndOfBibitem
\bibitem[Singh \latin{et~al.}(2016)Singh, Kaur, Sahoo, and Arora]{Singh2016}
Singh,~S.; Kaur,~K.; Sahoo,~B.~K.; Arora,~B. Comparing Magic Wavelengths for
  the Transitions of {{Cs}} Using Circularly and Linearly Polarized Light.
  \emph{J. Phys. B At. Mol. Opt. Phys.} \textbf{2016}, \emph{49}, 145005\relax
\mciteBstWouldAddEndPuncttrue
\mciteSetBstMidEndSepPunct{\mcitedefaultmidpunct}
{\mcitedefaultendpunct}{\mcitedefaultseppunct}\relax
\EndOfBibitem
\bibitem[Singh \latin{et~al.}(2016)Singh, Sahoo, and Arora]{Singh2016a}
Singh,~S.; Sahoo,~B.~K.; Arora,~B. Determination of Magic Wavelengths for the
  \$7s {\textasciicircum}\{2\}{{S}}\_\{1/2\}{\textbackslash}ensuremath\{-\}7p
  {\textasciicircum}\{2\}{{P}}\_\{3/2,1/2\}\$ Transitions in {{Fr}}.
  \emph{Phys. Rev. A} \textbf{2016}, \emph{94}, 023418\relax
\mciteBstWouldAddEndPuncttrue
\mciteSetBstMidEndSepPunct{\mcitedefaultmidpunct}
{\mcitedefaultendpunct}{\mcitedefaultseppunct}\relax
\EndOfBibitem
\bibitem[Aoki \latin{et~al.}(2021)Aoki, Sreekantham, Sahoo, Arora, Kastberg,
  Sato, Ikeda, Okamoto, Torii, Hayamizu, Nakamura, Nagase, Ohtsuka, Nagahama,
  Ozawa, Sato, Nakashita, Yamane, Tanaka, Harada, Kawamura, Inoue, Uchiyama,
  Hatakeyama, Takamine, Ueno, Ichikawa, Matsuda, Haba, and Sakemi]{Aoki2021}
Aoki,~T.; Sreekantham,~R.; Sahoo,~B.~K.; Arora,~B.; Kastberg,~A.; Sato,~T.;
  Ikeda,~H.; Okamoto,~N.; Torii,~Y.; Hayamizu,~T.; Nakamura,~K.; Nagase,~S.;
  Ohtsuka,~M.; Nagahama,~H.; Ozawa,~N.; Sato,~M.; Nakashita,~T.; Yamane,~K.;
  Tanaka,~K.~S.; Harada,~K.; Kawamura,~H.; Inoue,~T.; Uchiyama,~A.;
  Hatakeyama,~A.; Takamine,~A.; Ueno,~H.; Ichikawa,~Y.; Matsuda,~Y.; Haba,~H.;
  Sakemi,~Y. Quantum Sensing of the Electron Electric Dipole Moment Using
  Ultracold Entangled {{Fr}} Atoms. \emph{Quantum Sci. Technol.} \textbf{2021},
  \emph{6}, 044008\relax
\mciteBstWouldAddEndPuncttrue
\mciteSetBstMidEndSepPunct{\mcitedefaultmidpunct}
{\mcitedefaultendpunct}{\mcitedefaultseppunct}\relax
\EndOfBibitem
\bibitem[Fleig(2005)]{Fleig2005}
Fleig,~T. Spin-Orbit-Resolved Static Polarizabilities of Group-13 Atoms:
  {{Four-component}} Relativistic Configuration Interaction and Coupled Cluster
  Calculations. \emph{Phys. Rev. A} \textbf{2005}, \emph{72}, 052506\relax
\mciteBstWouldAddEndPuncttrue
\mciteSetBstMidEndSepPunct{\mcitedefaultmidpunct}
{\mcitedefaultendpunct}{\mcitedefaultseppunct}\relax
\EndOfBibitem
\bibitem[Pershina \latin{et~al.}(2008)Pershina, Borschevsky, Eliav, and
  Kaldor]{Pershina2008}
Pershina,~V.; Borschevsky,~A.; Eliav,~E.; Kaldor,~U. Adsorption of Inert Gases
  Including Element 118 on Noble Metal and Inert Surfaces from Ab Initio
  {{Dirac}}\textendash{{Coulomb}} Atomic Calculations. \emph{J. Chem. Phys.}
  \textbf{2008}, \emph{129}, 144106\relax
\mciteBstWouldAddEndPuncttrue
\mciteSetBstMidEndSepPunct{\mcitedefaultmidpunct}
{\mcitedefaultendpunct}{\mcitedefaultseppunct}\relax
\EndOfBibitem
\bibitem[Dzuba and Flambaum(2016)Dzuba, and Flambaum]{Dzuba2016}
Dzuba,~V.~A.; Flambaum,~V.~V. Electron Structure of Superheavy Elements
  {{Uut}}, {{Fl}} and {{Uup}} ({{Z}}=113 to 115). \emph{Hyperfine Interact.}
  \textbf{2016}, \emph{237}, 160\relax
\mciteBstWouldAddEndPuncttrue
\mciteSetBstMidEndSepPunct{\mcitedefaultmidpunct}
{\mcitedefaultendpunct}{\mcitedefaultseppunct}\relax
\EndOfBibitem
\bibitem[Fleig and Sadlej(2002)Fleig, and Sadlej]{Fleig2002}
Fleig,~T.; Sadlej,~A.~J. Electric dipole polarizabilities of the halogen atoms
  in ${}^{2}{P}_{1/2}$ and ${}^{2}{P}_{3/2}$ states: Scalar relativistic and
  two-component configuration-interaction calculations. \emph{Phys. Rev. A}
  \textbf{2002}, \emph{65}, 032506\relax
\mciteBstWouldAddEndPuncttrue
\mciteSetBstMidEndSepPunct{\mcitedefaultmidpunct}
{\mcitedefaultendpunct}{\mcitedefaultseppunct}\relax
\EndOfBibitem
\bibitem[{de Farias}(2017)]{deFarias2017}
{de Farias},~R.~F. Estimation of Some Physical Properties for Tennessine and
  Tennessine Hydride ({{TsH}}). \emph{Chem. Phys. Lett.} \textbf{2017},
  \emph{667}, 1--3\relax
\mciteBstWouldAddEndPuncttrue
\mciteSetBstMidEndSepPunct{\mcitedefaultmidpunct}
{\mcitedefaultendpunct}{\mcitedefaultseppunct}\relax
\EndOfBibitem
\bibitem[Schwerdtfeger and Nagle(2023)Schwerdtfeger, and
  Nagle]{Schwerdtfeger2023}
Schwerdtfeger,~P.; Nagle,~J.~K. 2023 {Table} of static dipole polarizabilities
  of the neutral elements in the periodic table. 2023;
  \url{https://ctcp.massey.ac.nz/2023Tablepol.pdf}, Accessed on Jun 14,
  2024\relax
\mciteBstWouldAddEndPuncttrue
\mciteSetBstMidEndSepPunct{\mcitedefaultmidpunct}
{\mcitedefaultendpunct}{\mcitedefaultseppunct}\relax
\EndOfBibitem
\bibitem[Yu \latin{et~al.}(2015)Yu, Suo, Feng, Fan, and Liu]{Yu2015}
Yu,~Y.-m.; Suo,~B.-b.; Feng,~H.-h.; Fan,~H.; Liu,~W.-M. Finite-field
  calculation of static polarizabilities and hyperpolarizabilities of
  ${\text{In}}^{+}$ and Sr. \emph{Phys. Rev. A} \textbf{2015}, \emph{92},
  052515\relax
\mciteBstWouldAddEndPuncttrue
\mciteSetBstMidEndSepPunct{\mcitedefaultmidpunct}
{\mcitedefaultendpunct}{\mcitedefaultseppunct}\relax
\EndOfBibitem
\bibitem[Fleig(2012)]{Fleig2012}
Fleig,~T. Invited Review: {{Relativistic}} Wave-Function Based Electron
  Correlation Methods. \emph{Chem. Phys.} \textbf{2012}, \emph{395},
  2--15\relax
\mciteBstWouldAddEndPuncttrue
\mciteSetBstMidEndSepPunct{\mcitedefaultmidpunct}
{\mcitedefaultendpunct}{\mcitedefaultseppunct}\relax
\EndOfBibitem
\bibitem[Saue(2011)]{Saue2011}
Saue,~T. Relativistic {{Hamiltonians}} for {{Chemistry}}: {{A Primer}}.
  \emph{ChemPhysChem} \textbf{2011}, \emph{12}, 3077--3094\relax
\mciteBstWouldAddEndPuncttrue
\mciteSetBstMidEndSepPunct{\mcitedefaultmidpunct}
{\mcitedefaultendpunct}{\mcitedefaultseppunct}\relax
\EndOfBibitem
\bibitem[Douglas and Kroll(1974)Douglas, and Kroll]{Douglas1974}
Douglas,~M.; Kroll,~N.~M. Quantum Electrodynamical Corrections to the Fine
  Structure of Helium. \emph{Ann. Phys.} \textbf{1974}, \emph{82},
  89--155\relax
\mciteBstWouldAddEndPuncttrue
\mciteSetBstMidEndSepPunct{\mcitedefaultmidpunct}
{\mcitedefaultendpunct}{\mcitedefaultseppunct}\relax
\EndOfBibitem
\bibitem[Hess(1985)]{Hess1985}
Hess,~B.~A. Applicability of the No-Pair Equation with Free-Particle Projection
  Operators to Atomic and Molecular Structure Calculations. \emph{Phys. Rev. A}
  \textbf{1985}, \emph{32}, 756--763\relax
\mciteBstWouldAddEndPuncttrue
\mciteSetBstMidEndSepPunct{\mcitedefaultmidpunct}
{\mcitedefaultendpunct}{\mcitedefaultseppunct}\relax
\EndOfBibitem
\bibitem[Hess(1986)]{Hess1986}
Hess,~B.~A. Relativistic Electronic-Structure Calculations Employing a
  Two-Component No-Pair Formalism with External-Field Projection Operators.
  \emph{Phys. Rev. A} \textbf{1986}, \emph{33}, 3742--3748\relax
\mciteBstWouldAddEndPuncttrue
\mciteSetBstMidEndSepPunct{\mcitedefaultmidpunct}
{\mcitedefaultendpunct}{\mcitedefaultseppunct}\relax
\EndOfBibitem
\bibitem[Chang \latin{et~al.}(1986)Chang, Pelissier, and Durand]{Chang1986}
Chang,~C.; Pelissier,~M.; Durand,~P. Regular {{Two-Component Pauli-Like
  Effective Hamiltonians}} in {{Dirac Theory}}. \emph{Phys. Scr.}
  \textbf{1986}, \emph{34}, 394\relax
\mciteBstWouldAddEndPuncttrue
\mciteSetBstMidEndSepPunct{\mcitedefaultmidpunct}
{\mcitedefaultendpunct}{\mcitedefaultseppunct}\relax
\EndOfBibitem
\bibitem[{van Lenthe} \latin{et~al.}(1994){van Lenthe}, Baerends, and
  Snijders]{vanLenthe1994}
{van Lenthe},~E.; Baerends,~E.~J.; Snijders,~J.~G. Relativistic Total Energy
  Using Regular Approximations. \emph{J. Chem. Phys.} \textbf{1994},
  \emph{101}, 9783--9792\relax
\mciteBstWouldAddEndPuncttrue
\mciteSetBstMidEndSepPunct{\mcitedefaultmidpunct}
{\mcitedefaultendpunct}{\mcitedefaultseppunct}\relax
\EndOfBibitem
\bibitem[{van Lenthe} \latin{et~al.}(1996){van Lenthe}, Snijders, and
  Baerends]{vanLenthe1996}
{van Lenthe},~E.; Snijders,~J.~G.; Baerends,~E.~J. The Zero-order Regular
  Approximation for Relativistic Effects: {{The}} Effect of Spin\textendash
  Orbit Coupling in Closed Shell Molecules. \emph{J. Chem. Phys.}
  \textbf{1996}, \emph{105}, 6505--6516\relax
\mciteBstWouldAddEndPuncttrue
\mciteSetBstMidEndSepPunct{\mcitedefaultmidpunct}
{\mcitedefaultendpunct}{\mcitedefaultseppunct}\relax
\EndOfBibitem
\bibitem[Ilia{\v s} and Saue(2007)Ilia{\v s}, and Saue]{Ilias2007}
Ilia{\v s},~M.; Saue,~T. An Infinite-Order Two-Component Relativistic
  {{Hamiltonian}} by a Simple One-Step Transformation. \emph{J. Chem. Phys.}
  \textbf{2007}, \emph{126}, 064102\relax
\mciteBstWouldAddEndPuncttrue
\mciteSetBstMidEndSepPunct{\mcitedefaultmidpunct}
{\mcitedefaultendpunct}{\mcitedefaultseppunct}\relax
\EndOfBibitem
\bibitem[Saue \latin{et~al.}(2020)Saue, Bast, Gomes, Jensen, Visscher, Aucar,
  Di~Remigio, Dyall, Eliav, Fasshauer, Fleig, Halbert, Hedeg{\aa}rd,
  {Helmich-Paris}, Ilia{\v s}, Jacob, Knecht, Laerdahl, Vidal, Nayak,
  Olejniczak, Olsen, Pernpointner, Senjean, Shee, Sunaga, and {van
  Stralen}]{Saue2020}
Saue,~T.; Bast,~R.; Gomes,~A. S.~P.; Jensen,~H. J.~A.; Visscher,~L.;
  Aucar,~I.~A.; Di~Remigio,~R.; Dyall,~K.~G.; Eliav,~E.; Fasshauer,~E.;
  Fleig,~T.; Halbert,~L.; Hedeg{\aa}rd,~E.~D.; {Helmich-Paris},~B.; Ilia{\v
  s},~M.; Jacob,~C.~R.; Knecht,~S.; Laerdahl,~J.~K.; Vidal,~M.~L.;
  Nayak,~M.~K.; Olejniczak,~M.; Olsen,~J. M.~H.; Pernpointner,~M.; Senjean,~B.;
  Shee,~A.; Sunaga,~A.; {van Stralen},~J. N.~P. The {{DIRAC}} Code for
  Relativistic Molecular Calculations. \emph{J. Chem. Phys.} \textbf{2020},
  \emph{152}, 204104\relax
\mciteBstWouldAddEndPuncttrue
\mciteSetBstMidEndSepPunct{\mcitedefaultmidpunct}
{\mcitedefaultendpunct}{\mcitedefaultseppunct}\relax
\EndOfBibitem
\bibitem[{van Stralen} \latin{et~al.}(2005){van Stralen}, Visscher, Larsen, and
  Jensen]{vanStralen2005a}
{van Stralen},~J. N.~P.; Visscher,~L.; Larsen,~C.~V.; Jensen,~H. J.~A.
  First-Order {{MP2}} Molecular Properties in a Relativistic Framework.
  \emph{Chem. Phys.} \textbf{2005}, \emph{311}, 81--95\relax
\mciteBstWouldAddEndPuncttrue
\mciteSetBstMidEndSepPunct{\mcitedefaultmidpunct}
{\mcitedefaultendpunct}{\mcitedefaultseppunct}\relax
\EndOfBibitem
\bibitem[Visscher \latin{et~al.}(1996)Visscher, Lee, and Dyall]{Visscher1996}
Visscher,~L.; Lee,~T.~J.; Dyall,~K.~G. Formulation and Implementation of a
  Relativistic Unrestricted Coupled-cluster Method Including Noniterative
  Connected Triples. \emph{J. Chem. Phys.} \textbf{1996}, \emph{105},
  8769--8776\relax
\mciteBstWouldAddEndPuncttrue
\mciteSetBstMidEndSepPunct{\mcitedefaultmidpunct}
{\mcitedefaultendpunct}{\mcitedefaultseppunct}\relax
\EndOfBibitem
\bibitem[Fleig \latin{et~al.}(2003)Fleig, Olsen, and Visscher]{Fleig2003}
Fleig,~T.; Olsen,~J.; Visscher,~L. The Generalized Active Space Concept for the
  Relativistic Treatment of Electron Correlation. {{II}}. {{Large-scale}}
  Configuration Interaction Implementation Based on Relativistic 2- and
  4-Spinors and Its Application. \emph{J. Chem. Phys.} \textbf{2003},
  \emph{119}, 2963--2971\relax
\mciteBstWouldAddEndPuncttrue
\mciteSetBstMidEndSepPunct{\mcitedefaultmidpunct}
{\mcitedefaultendpunct}{\mcitedefaultseppunct}\relax
\EndOfBibitem
\bibitem[Fleig \latin{et~al.}(2006)Fleig, Jensen, Olsen, and
  Visscher]{Fleig2006}
Fleig,~T.; Jensen,~H. J.~A.; Olsen,~J.; Visscher,~L. The Generalized Active
  Space Concept for the Relativistic Treatment of Electron Correlation.
  {{III}}. {{Large-scale}} Configuration Interaction and Multiconfiguration
  Self-Consistent-Field Four-Component Methods with Application to {{UO2}}.
  \emph{J. Chem. Phys.} \textbf{2006}, \emph{124}, 104106\relax
\mciteBstWouldAddEndPuncttrue
\mciteSetBstMidEndSepPunct{\mcitedefaultmidpunct}
{\mcitedefaultendpunct}{\mcitedefaultseppunct}\relax
\EndOfBibitem
\bibitem[Knecht \latin{et~al.}(2008)Knecht, Jensen, and Fleig]{Knecht2008}
Knecht,~S.; Jensen,~H. J.~A.; Fleig,~T. Large-Scale Parallel Configuration
  Interaction. {{I}}. {{Nonrelativistic}} and Scalar-Relativistic General
  Active Space Implementation with Application to ({{Rb}}\textendash{{Ba}})+.
  \emph{J. Chem. Phys.} \textbf{2008}, \emph{128}, 014108\relax
\mciteBstWouldAddEndPuncttrue
\mciteSetBstMidEndSepPunct{\mcitedefaultmidpunct}
{\mcitedefaultendpunct}{\mcitedefaultseppunct}\relax
\EndOfBibitem
\bibitem[Knecht \latin{et~al.}(2010)Knecht, Jensen, and Fleig]{Knecht2010}
Knecht,~S.; Jensen,~H. J.~A.; Fleig,~T. Large-Scale Parallel Configuration
  Interaction. {{II}}. {{Two-}} and Four-Component Double-Group General Active
  Space Implementation with Application to {{BiH}}. \emph{J. Chem. Phys.}
  \textbf{2010}, \emph{132}, 014108\relax
\mciteBstWouldAddEndPuncttrue
\mciteSetBstMidEndSepPunct{\mcitedefaultmidpunct}
{\mcitedefaultendpunct}{\mcitedefaultseppunct}\relax
\EndOfBibitem
\bibitem[Das and Thakkar(1998)Das, and Thakkar]{Das1998}
Das,~A.~K.; Thakkar,~A.~J. Static Response Properties of Second-Period Atoms:
  Coupled Cluster Calculations. \emph{J. Phys. B At. Mol. Opt. Phys.}
  \textbf{1998}, \emph{31}, 2215\relax
\mciteBstWouldAddEndPuncttrue
\mciteSetBstMidEndSepPunct{\mcitedefaultmidpunct}
{\mcitedefaultendpunct}{\mcitedefaultseppunct}\relax
\EndOfBibitem
\bibitem[Kassimi and Thakkar(1994)Kassimi, and Thakkar]{Kassimi1994}
Kassimi,~N. E.-B.; Thakkar,~A.~J. Static hyperpolarizability of atomic lithium.
  \emph{Physical Review A} \textbf{1994}, \emph{50}, 2948--2952, Publisher:
  American Physical Society\relax
\mciteBstWouldAddEndPuncttrue
\mciteSetBstMidEndSepPunct{\mcitedefaultmidpunct}
{\mcitedefaultendpunct}{\mcitedefaultseppunct}\relax
\EndOfBibitem
\bibitem[Kállay \latin{et~al.}(2011)Kállay, Nataraj, Sahoo, Das, and
  Visscher]{Kallay2011}
Kállay,~M.; Nataraj,~H.~S.; Sahoo,~B.~K.; Das,~B.~P.; Visscher,~L.
  Relativistic general-order coupled-cluster method for high-precision
  calculations: {Application} to the {Al}\$\{\}{\textasciicircum}\{+\}\$ atomic
  clock. \emph{Phys. Rev. A} \textbf{2011}, \emph{83}, 030503\relax
\mciteBstWouldAddEndPuncttrue
\mciteSetBstMidEndSepPunct{\mcitedefaultmidpunct}
{\mcitedefaultendpunct}{\mcitedefaultseppunct}\relax
\EndOfBibitem
\bibitem[Irikura(2021)]{Irikura2021}
Irikura,~K.~K. Polarizability of atomic {Pt}, {Pt}+, and {Pt}-. \emph{J. Chem.
  Phys.} \textbf{2021}, \emph{154}, 174302\relax
\mciteBstWouldAddEndPuncttrue
\mciteSetBstMidEndSepPunct{\mcitedefaultmidpunct}
{\mcitedefaultendpunct}{\mcitedefaultseppunct}\relax
\EndOfBibitem
\bibitem[dir()]{dirac_basis}
Pick the right basis for your calculation.
  \url{https://www.diracprogram.org/doc/release-24/molecule_and_basis/basis.html},
  Accessed: 2024-06-01\relax
\mciteBstWouldAddEndPuncttrue
\mciteSetBstMidEndSepPunct{\mcitedefaultmidpunct}
{\mcitedefaultendpunct}{\mcitedefaultseppunct}\relax
\EndOfBibitem
\bibitem[Williams(2016)]{Williams2016}
Williams,~J.~H. \emph{Quantifying {Measurement}: {The} tyranny of numbers};
  Morgan \& Claypool Publishers, 2016\relax
\mciteBstWouldAddEndPuncttrue
\mciteSetBstMidEndSepPunct{\mcitedefaultmidpunct}
{\mcitedefaultendpunct}{\mcitedefaultseppunct}\relax
\EndOfBibitem
\bibitem[pyd()]{pydirac}
pydirac 2024.7.8. \url{https://pypi.org/project/pydirac/}, Accessed:
  2024-07-08\relax
\mciteBstWouldAddEndPuncttrue
\mciteSetBstMidEndSepPunct{\mcitedefaultmidpunct}
{\mcitedefaultendpunct}{\mcitedefaultseppunct}\relax
\EndOfBibitem
\bibitem[Dyall(2002)]{Dyall2002}
Dyall,~K.~G. Relativistic and Nonrelativistic Finite Nucleus Optimized
  Triple-Zeta Basis Sets for the 4p, 5p and 6p Elements. \emph{Theor. Chem.
  Acc.} \textbf{2002}, \emph{108}, 335--340\relax
\mciteBstWouldAddEndPuncttrue
\mciteSetBstMidEndSepPunct{\mcitedefaultmidpunct}
{\mcitedefaultendpunct}{\mcitedefaultseppunct}\relax
\EndOfBibitem
\bibitem[Dyall(2004)]{Dyall2004}
Dyall,~K.~G. Relativistic Double-Zeta, Triple-Zeta, and Quadruple-Zeta Basis
  Sets for the 5d Elements {{Hf}}\textendash{{Hg}}. \emph{Theor. Chem. Acc.}
  \textbf{2004}, \emph{112}, 403--409\relax
\mciteBstWouldAddEndPuncttrue
\mciteSetBstMidEndSepPunct{\mcitedefaultmidpunct}
{\mcitedefaultendpunct}{\mcitedefaultseppunct}\relax
\EndOfBibitem
\bibitem[Dyall(2006)]{Dyall2006}
Dyall,~K.~G. Relativistic {{Quadruple-Zeta}} and {{Revised Triple-Zeta}} and
  {{Double-Zeta Basis Sets}} for the 4p, 5p, and 6p {{Elements}}. \emph{Theor.
  Chem. Acc.} \textbf{2006}, \emph{115}, 441--447\relax
\mciteBstWouldAddEndPuncttrue
\mciteSetBstMidEndSepPunct{\mcitedefaultmidpunct}
{\mcitedefaultendpunct}{\mcitedefaultseppunct}\relax
\EndOfBibitem
\bibitem[Dyall(2007)]{Dyall2007}
Dyall,~K.~G. Relativistic Double-Zeta, Triple-Zeta, and Quadruple-Zeta Basis
  Sets for the 4d Elements {{Y}}\textendash{{Cd}}. \emph{Theor. Chem. Acc.}
  \textbf{2007}, \emph{117}, 483--489\relax
\mciteBstWouldAddEndPuncttrue
\mciteSetBstMidEndSepPunct{\mcitedefaultmidpunct}
{\mcitedefaultendpunct}{\mcitedefaultseppunct}\relax
\EndOfBibitem
\bibitem[Dyall(2009)]{Dyall2009}
Dyall,~K.~G. Relativistic {{Double-Zeta}}, {{Triple-Zeta}}, and
  {{Quadruple-Zeta Basis Sets}} for the 4s, 5s, 6s, and 7s {{Elements}}.
  \emph{J. Phys. Chem. A} \textbf{2009}, \emph{113}, 12638--12644\relax
\mciteBstWouldAddEndPuncttrue
\mciteSetBstMidEndSepPunct{\mcitedefaultmidpunct}
{\mcitedefaultendpunct}{\mcitedefaultseppunct}\relax
\EndOfBibitem
\bibitem[Dyall and Gomes(2010)Dyall, and Gomes]{Dyall2010}
Dyall,~K.~G.; Gomes,~A. S.~P. Revised Relativistic Basis Sets for the 5d
  Elements {{Hf}}\textendash{{Hg}}. \emph{Theor. Chem. Acc.} \textbf{2010},
  \emph{125}, 97--100\relax
\mciteBstWouldAddEndPuncttrue
\mciteSetBstMidEndSepPunct{\mcitedefaultmidpunct}
{\mcitedefaultendpunct}{\mcitedefaultseppunct}\relax
\EndOfBibitem
\bibitem[Dyall(2011)]{Dyall2011}
Dyall,~K.~G. Relativistic Double-Zeta, Triple-Zeta, and Quadruple-Zeta Basis
  Sets for the 6d Elements {{Rf}}\textendash{{Cn}}. \emph{Theor. Chem. Acc.}
  \textbf{2011}, \emph{129}, 603--613\relax
\mciteBstWouldAddEndPuncttrue
\mciteSetBstMidEndSepPunct{\mcitedefaultmidpunct}
{\mcitedefaultendpunct}{\mcitedefaultseppunct}\relax
\EndOfBibitem
\bibitem[Dyall(2016)]{Dyall2016}
Dyall,~K.~G. Relativistic Double-Zeta, Triple-Zeta, and Quadruple-Zeta Basis
  Sets for the Light Elements {{H}}\textendash{{Ar}}. \emph{Theor. Chem. Acc.}
  \textbf{2016}, \emph{135}, 128\relax
\mciteBstWouldAddEndPuncttrue
\mciteSetBstMidEndSepPunct{\mcitedefaultmidpunct}
{\mcitedefaultendpunct}{\mcitedefaultseppunct}\relax
\EndOfBibitem
\bibitem[Roos \latin{et~al.}(2004)Roos, Lindh, Malmqvist, Veryazov, and
  Widmark]{Roos2004}
Roos,~B.~O.; Lindh,~R.; Malmqvist,~P.-{\AA}.; Veryazov,~V.; Widmark,~P.-O. Main
  {{Group Atoms}} and {{Dimers Studied}} with a {{New Relativistic ANO Basis
  Set}}. \emph{J. Phys. Chem. A} \textbf{2004}, \emph{108}, 2851--2858\relax
\mciteBstWouldAddEndPuncttrue
\mciteSetBstMidEndSepPunct{\mcitedefaultmidpunct}
{\mcitedefaultendpunct}{\mcitedefaultseppunct}\relax
\EndOfBibitem
\bibitem[Roos \latin{et~al.}(2005)Roos, Lindh, Malmqvist, Veryazov, and
  Widmark]{Roos2005}
Roos,~B.~O.; Lindh,~R.; Malmqvist,~P.-{\AA}.~{\AA}.; Veryazov,~V.;
  Widmark,~P.-O.~O. New {{Relativistic ANO Basis Sets}} for {{Transition Metal
  Atoms}}. \emph{J. Phys. Chem. A} \textbf{2005}, \emph{109}, 6575--6579\relax
\mciteBstWouldAddEndPuncttrue
\mciteSetBstMidEndSepPunct{\mcitedefaultmidpunct}
{\mcitedefaultendpunct}{\mcitedefaultseppunct}\relax
\EndOfBibitem
\bibitem[DIR()]{DIRAC18}
DIRAC18. \url{https://doi.org/10.5281/zenodo.2253986}, Accessed on Jun 7,
  2021\relax
\mciteBstWouldAddEndPuncttrue
\mciteSetBstMidEndSepPunct{\mcitedefaultmidpunct}
{\mcitedefaultendpunct}{\mcitedefaultseppunct}\relax
\EndOfBibitem
\bibitem[Molof \latin{et~al.}(1974)Molof, Schwartz, Miller, and
  Bederson]{Molof1974}
Molof,~R.~W.; Schwartz,~H.~L.; Miller,~T.~M.; Bederson,~B. Measurements of
  Electric Dipole Polarizabilities of the Alkali-Metal Atoms and the Metastable
  Noble-Gas Atoms. \emph{Phys. Rev. A} \textbf{1974}, \emph{10},
  1131--1140\relax
\mciteBstWouldAddEndPuncttrue
\mciteSetBstMidEndSepPunct{\mcitedefaultmidpunct}
{\mcitedefaultendpunct}{\mcitedefaultseppunct}\relax
\EndOfBibitem
\bibitem[Miffre \latin{et~al.}(2006)Miffre, Jacquey, B{\"u}chner, Tr{\'e}nec,
  and Vigu{\'e}]{Miffre2006a}
Miffre,~A.; Jacquey,~M.; B{\"u}chner,~M.; Tr{\'e}nec,~G.; Vigu{\'e},~J.
  Measurement of the Electric Polarizability of Lithium by Atom Interferometry.
  \emph{Phys. Rev. A} \textbf{2006}, \emph{73}, 011603\relax
\mciteBstWouldAddEndPuncttrue
\mciteSetBstMidEndSepPunct{\mcitedefaultmidpunct}
{\mcitedefaultendpunct}{\mcitedefaultseppunct}\relax
\EndOfBibitem
\bibitem[Puchalski \latin{et~al.}(2011)Puchalski, K{\k e}dziera, and
  Pachucki]{Puchalski2011}
Puchalski,~M.; K{\k e}dziera,~D.; Pachucki,~K. Lithium Electric Dipole
  Polarizability. \emph{Phys. Rev. A} \textbf{2011}, \emph{84}, 052518\relax
\mciteBstWouldAddEndPuncttrue
\mciteSetBstMidEndSepPunct{\mcitedefaultmidpunct}
{\mcitedefaultendpunct}{\mcitedefaultseppunct}\relax
\EndOfBibitem
\bibitem[Ma \latin{et~al.}(2015)Ma, Indergaard, Zhang, Larkin, Moro, and {de
  Heer}]{Ma2015}
Ma,~L.; Indergaard,~J.; Zhang,~B.; Larkin,~I.; Moro,~R.; {de Heer},~W.~A.
  Measured Atomic Ground-State Polarizabilities of 35 Metallic Elements.
  \emph{Phys. Rev. A} \textbf{2015}, \emph{91}, 010501\relax
\mciteBstWouldAddEndPuncttrue
\mciteSetBstMidEndSepPunct{\mcitedefaultmidpunct}
{\mcitedefaultendpunct}{\mcitedefaultseppunct}\relax
\EndOfBibitem
\bibitem[Maroulis(2006)]{Maroulis2006}
Maroulis,~G. \emph{Atoms, {{Molecules And Clusters In Electric Fields}}:
  {{Theoretical Approaches To The Calculation Of Electric Polarizability}}};
  World Scientific, 2006\relax
\mciteBstWouldAddEndPuncttrue
\mciteSetBstMidEndSepPunct{\mcitedefaultmidpunct}
{\mcitedefaultendpunct}{\mcitedefaultseppunct}\relax
\EndOfBibitem
\bibitem[Thakkar and Lupinetti(2005)Thakkar, and Lupinetti]{Thakkar2005}
Thakkar,~A.~J.; Lupinetti,~C. The Polarizability of Sodium: Theory and
  Experiment Reconciled. \emph{Chem. Phys. Lett.} \textbf{2005}, \emph{402},
  270--273\relax
\mciteBstWouldAddEndPuncttrue
\mciteSetBstMidEndSepPunct{\mcitedefaultmidpunct}
{\mcitedefaultendpunct}{\mcitedefaultseppunct}\relax
\EndOfBibitem
\bibitem[Lim \latin{et~al.}(1999)Lim, Pernpointner, Seth, Laerdahl,
  Schwerdtfeger, Neogrady, and Urban]{Lim1999}
Lim,~I.~S.; Pernpointner,~M.; Seth,~M.; Laerdahl,~J.~K.; Schwerdtfeger,~P.;
  Neogrady,~P.; Urban,~M. Relativistic Coupled-Cluster Static Dipole
  Polarizabilities of the Alkali Metals from {{Li}} to Element 119. \emph{Phys.
  Rev. A} \textbf{1999}, \emph{60}, 2822--2828\relax
\mciteBstWouldAddEndPuncttrue
\mciteSetBstMidEndSepPunct{\mcitedefaultmidpunct}
{\mcitedefaultendpunct}{\mcitedefaultseppunct}\relax
\EndOfBibitem
\bibitem[{\'S}mia{\l}kowski and Tomza(2021){\'S}mia{\l}kowski, and
  Tomza]{Smialkowski2021}
{\'S}mia{\l}kowski,~M.; Tomza,~M. Highly Polar Molecules Consisting of a Copper
  or Silver Atom Interacting with an Alkali-Metal or Alkaline-Earth-Metal Atom.
  \emph{Phys. Rev. A} \textbf{2021}, \emph{103}, 022802\relax
\mciteBstWouldAddEndPuncttrue
\mciteSetBstMidEndSepPunct{\mcitedefaultmidpunct}
{\mcitedefaultendpunct}{\mcitedefaultseppunct}\relax
\EndOfBibitem
\bibitem[Gregoire \latin{et~al.}(2015)Gregoire, Hromada, Holmgren, Trubko, and
  Cronin]{Gregoire2015}
Gregoire,~M.~D.; Hromada,~I.; Holmgren,~W.~F.; Trubko,~R.; Cronin,~A.~D.
  Measurements of the Ground-State Polarizabilities of {{Cs}}, {{Rb}}, and
  {{K}} Using Atom Interferometry. \emph{Phys. Rev. A} \textbf{2015},
  \emph{92}, 052513\relax
\mciteBstWouldAddEndPuncttrue
\mciteSetBstMidEndSepPunct{\mcitedefaultmidpunct}
{\mcitedefaultendpunct}{\mcitedefaultseppunct}\relax
\EndOfBibitem
\bibitem[Gregoire \latin{et~al.}(2016)Gregoire, Brooks, Trubko, and
  Cronin]{Gregoire2016}
Gregoire,~M.~D.; Brooks,~N.; Trubko,~R.; Cronin,~A.~D. Analysis of
  {{Polarizability Measurements Made}} with {{Atom Interferometry}}.
  \emph{Atoms} \textbf{2016}, \emph{4}, 21\relax
\mciteBstWouldAddEndPuncttrue
\mciteSetBstMidEndSepPunct{\mcitedefaultmidpunct}
{\mcitedefaultendpunct}{\mcitedefaultseppunct}\relax
\EndOfBibitem
\bibitem[Faegri~Jr(2001)]{FaegriJr2001}
Faegri~Jr,~K. Relativistic {Gaussian} basis sets for the elements {K} –
  {Uuo}. \emph{Theor. Chem. Acc.} \textbf{2001}, \emph{105}, 252--258\relax
\mciteBstWouldAddEndPuncttrue
\mciteSetBstMidEndSepPunct{\mcitedefaultmidpunct}
{\mcitedefaultendpunct}{\mcitedefaultseppunct}\relax
\EndOfBibitem
\bibitem[Lim \latin{et~al.}(2005)Lim, Schwerdtfeger, Metz, and Stoll]{Lim2005}
Lim,~I.~S.; Schwerdtfeger,~P.; Metz,~B.; Stoll,~H. All-Electron and
  Relativistic Pseudopotential Studies for the Group 1 Element Polarizabilities
  from {{K}} to Element 119. \emph{J. Chem. Phys.} \textbf{2005}, \emph{122},
  104103\relax
\mciteBstWouldAddEndPuncttrue
\mciteSetBstMidEndSepPunct{\mcitedefaultmidpunct}
{\mcitedefaultendpunct}{\mcitedefaultseppunct}\relax
\EndOfBibitem
\bibitem[Jiang \latin{et~al.}(2015)Jiang, Mitroy, Cheng, and
  Bromley]{Jiang2015a}
Jiang,~J.; Mitroy,~J.; Cheng,~Y.; Bromley,~M. W.~J. Effective Oscillator
  Strength Distributions of Spherically Symmetric Atoms for Calculating
  Polarizabilities and Long-Range Atom--Atom Interactions. \emph{At. Data Nucl.
  Data Tables} \textbf{2015}, \emph{101}, 158--186\relax
\mciteBstWouldAddEndPuncttrue
\mciteSetBstMidEndSepPunct{\mcitedefaultmidpunct}
{\mcitedefaultendpunct}{\mcitedefaultseppunct}\relax
\EndOfBibitem
\bibitem[Cheng \latin{et~al.}(2013)Cheng, Jiang, and Mitroy]{Cheng2013}
Cheng,~Y.; Jiang,~J.; Mitroy,~J. Tune-out Wavelengths for the
  Alkaline-Earth-Metal Atoms. \emph{Phys. Rev. A} \textbf{2013}, \emph{88},
  022511\relax
\mciteBstWouldAddEndPuncttrue
\mciteSetBstMidEndSepPunct{\mcitedefaultmidpunct}
{\mcitedefaultendpunct}{\mcitedefaultseppunct}\relax
\EndOfBibitem
\bibitem[Wu \latin{et~al.}(2023)Wu, Wang, Wang, Jiang, and Dong]{Wu2023}
Wu,~L.; Wang,~X.; Wang,~T.; Jiang,~J.; Dong,~C. Be Optical Lattice Clocks with
  the Fractional {{Stark}} Shift up to the Level of 10-19. \emph{New J. Phys.}
  \textbf{2023}, \emph{25}, 043011\relax
\mciteBstWouldAddEndPuncttrue
\mciteSetBstMidEndSepPunct{\mcitedefaultmidpunct}
{\mcitedefaultendpunct}{\mcitedefaultseppunct}\relax
\EndOfBibitem
\bibitem[Porsev and Derevianko(2006)Porsev, and Derevianko]{Porsev2006}
Porsev,~S.~G.; Derevianko,~A. High-Accuracy Calculations of Dipole, Quadrupole,
  and Octupole Electric Dynamic Polarizabilities and van Der {{Waals}}
  Coefficients {{C6}}, {{C8}}, and {{C10}} for Alkaline-Earth Dimers. \emph{J.
  Exp. Theor. Phys.} \textbf{2006}, \emph{102}, 195--205\relax
\mciteBstWouldAddEndPuncttrue
\mciteSetBstMidEndSepPunct{\mcitedefaultmidpunct}
{\mcitedefaultendpunct}{\mcitedefaultseppunct}\relax
\EndOfBibitem
\bibitem[Derevianko \latin{et~al.}(2010)Derevianko, Porsev, and
  Babb]{Derevianko2010}
Derevianko,~A.; Porsev,~S.~G.; Babb,~J.~F. Electric dipole polarizabilities at
  imaginary frequencies for hydrogen, the alkali–metal, alkaline–earth, and
  noble gas atoms. \emph{Atomic Data and Nuclear Data Tables} \textbf{2010},
  \emph{96}, 323--331\relax
\mciteBstWouldAddEndPuncttrue
\mciteSetBstMidEndSepPunct{\mcitedefaultmidpunct}
{\mcitedefaultendpunct}{\mcitedefaultseppunct}\relax
\EndOfBibitem
\bibitem[Thierfelder \latin{et~al.}(2008)Thierfelder, Assadollahzadeh,
  Schwerdtfeger, Sch{\"a}fer, and Sch{\"a}fer]{Thierfelder2008}
Thierfelder,~C.; Assadollahzadeh,~B.; Schwerdtfeger,~P.; Sch{\"a}fer,~S.;
  Sch{\"a}fer,~R. Relativistic and Electron Correlation Effects in Static
  Dipole Polarizabilities for the Group-14 Elements from Carbon to Element
  $Z=114$: Theory and Experiment. \emph{Phys. Rev. A} \textbf{2008}, \emph{78},
  052506\relax
\mciteBstWouldAddEndPuncttrue
\mciteSetBstMidEndSepPunct{\mcitedefaultmidpunct}
{\mcitedefaultendpunct}{\mcitedefaultseppunct}\relax
\EndOfBibitem
\bibitem[Canal~Neto \latin{et~al.}(2021)Canal~Neto, Ferreira, Jorge, and {de
  Oliveira}]{CanalNeto2021}
Canal~Neto,~A.; Ferreira,~I.~B.; Jorge,~F.~E.; {de Oliveira},~A.~Z.
  All-Electron Triple Zeta Basis Sets for {{ZORA}} Calculations:
  {{Application}} in Studies of Atoms and Molecules. \emph{Chemical Physics
  Letters} \textbf{2021}, \emph{771}, 138548\relax
\mciteBstWouldAddEndPuncttrue
\mciteSetBstMidEndSepPunct{\mcitedefaultmidpunct}
{\mcitedefaultendpunct}{\mcitedefaultseppunct}\relax
\EndOfBibitem
\bibitem[A.~Manz \latin{et~al.}(2019)A.~Manz, Chen, J.~Cole, Gabaldon~Limas,
  and Fiszbein]{A.Manz2019}
A.~Manz,~T.; Chen,~T.; J.~Cole,~D.; Gabaldon~Limas,~N.; Fiszbein,~B. New
  Scaling Relations to Compute Atom-in-Material Polarizabilities and Dispersion
  Coefficients: Part 1. {{Theory}} and Accuracy. \emph{RSC Adv.} \textbf{2019},
  \emph{9}, 19297--19324\relax
\mciteBstWouldAddEndPuncttrue
\mciteSetBstMidEndSepPunct{\mcitedefaultmidpunct}
{\mcitedefaultendpunct}{\mcitedefaultseppunct}\relax
\EndOfBibitem
\bibitem[Wang \latin{et~al.}(2021)Wang, Wang, Fan, Zhao, Miao, Yin, Moro, and
  Ma]{Wang2021}
Wang,~K.; Wang,~X.; Fan,~Z.; Zhao,~H.-Y.; Miao,~L.; Yin,~G.-J.; Moro,~R.;
  Ma,~L. Static Dipole Polarizabilities of Atoms and Ions from {{Z}} = 1 to 20
  Calculated within a Single Theoretical Scheme. \emph{Eur. Phys. J. D}
  \textbf{2021}, \emph{75}, 1--11\relax
\mciteBstWouldAddEndPuncttrue
\mciteSetBstMidEndSepPunct{\mcitedefaultmidpunct}
{\mcitedefaultendpunct}{\mcitedefaultseppunct}\relax
\EndOfBibitem
\bibitem[{\'E}hn and {\v C}ernu{\v s}{\'a}k(2021){\'E}hn, and {\v C}ernu{\v
  s}{\'a}k]{Ehn2021}
{\'E}hn,~L.; {\v C}ernu{\v s}{\'a}k,~I. Atomic and Ionic Polarizabilities of
  {{B}}, {{C}}, {{N}}, {{O}}, and {{F}}. \emph{Int. J. Quantum Chem.}
  \textbf{2021}, \emph{121}, e26467\relax
\mciteBstWouldAddEndPuncttrue
\mciteSetBstMidEndSepPunct{\mcitedefaultmidpunct}
{\mcitedefaultendpunct}{\mcitedefaultseppunct}\relax
\EndOfBibitem
\bibitem[Doolen and Liberman(1987)Doolen, and Liberman]{Doolen1987}
Doolen,~G.; Liberman,~D.~A. Calculations of Photoabsorption by Atoms Using a
  Linear Response Method. \emph{Phys. Scr.} \textbf{1987}, \emph{36}, 77\relax
\mciteBstWouldAddEndPuncttrue
\mciteSetBstMidEndSepPunct{\mcitedefaultmidpunct}
{\mcitedefaultendpunct}{\mcitedefaultseppunct}\relax
\EndOfBibitem
\bibitem[Huot and Bose(1991)Huot, and Bose]{Huot1991}
Huot,~J.; Bose,~T.~K. Experimental Determination of the Dielectric Virial
  Coefficients of Atomic Gases as a Function of Temperature. \emph{The Journal
  of Chemical Physics} \textbf{1991}, \emph{95}, 2683--2687\relax
\mciteBstWouldAddEndPuncttrue
\mciteSetBstMidEndSepPunct{\mcitedefaultmidpunct}
{\mcitedefaultendpunct}{\mcitedefaultseppunct}\relax
\EndOfBibitem
\bibitem[Dalgarno \latin{et~al.}(1997)Dalgarno, Kingston, and
  Bates]{Dalgarno1997b}
Dalgarno,~A.; Kingston,~A.~E.; Bates,~D.~R. The Refractive Indices and
  {{Verdet}} Constants of the Inert Gases. \emph{Proc. R. Soc. Lond. Ser. Math.
  Phys. Sci.} \textbf{1997}, \emph{259}, 424--431\relax
\mciteBstWouldAddEndPuncttrue
\mciteSetBstMidEndSepPunct{\mcitedefaultmidpunct}
{\mcitedefaultendpunct}{\mcitedefaultseppunct}\relax
\EndOfBibitem
\bibitem[Gaiser and Fellmuth(2018)Gaiser, and Fellmuth]{Gaiser2018}
Gaiser,~C.; Fellmuth,~B. Polarizability of {{Helium}}, {{Neon}}, and {{Argon}}:
  {{New Perspectives}} for {{Gas Metrology}}. \emph{Phys. Rev. Lett.}
  \textbf{2018}, \emph{120}, 123203\relax
\mciteBstWouldAddEndPuncttrue
\mciteSetBstMidEndSepPunct{\mcitedefaultmidpunct}
{\mcitedefaultendpunct}{\mcitedefaultseppunct}\relax
\EndOfBibitem
\bibitem[Sold{\'a}n \latin{et~al.}(2001)Sold{\'a}n, Lee, and
  Wright]{Soldán2001a}
Sold{\'a}n,~P.; Lee,~E. P.~F.; Wright,~T.~G. Static Dipole Polarizabilities
  ({$\alpha$}) and Static Second Hyperpolarizabilities ({$\gamma$}) of the Rare
  Gas Atoms ({{He}}--{{Rn}}). \emph{Phys. Chem. Chem. Phys.} \textbf{2001},
  \emph{3}, 4661--4666\relax
\mciteBstWouldAddEndPuncttrue
\mciteSetBstMidEndSepPunct{\mcitedefaultmidpunct}
{\mcitedefaultendpunct}{\mcitedefaultseppunct}\relax
\EndOfBibitem
\bibitem[Lesiuk \latin{et~al.}(2020)Lesiuk, Przybytek, and
  Jeziorski]{Lesiuk2020}
Lesiuk,~M.; Przybytek,~M.; Jeziorski,~B. Theoretical Determination of
  Polarizability and Magnetic Susceptibility of Neon. \emph{Phys. Rev. A}
  \textbf{2020}, \emph{102}, 052816\relax
\mciteBstWouldAddEndPuncttrue
\mciteSetBstMidEndSepPunct{\mcitedefaultmidpunct}
{\mcitedefaultendpunct}{\mcitedefaultseppunct}\relax
\EndOfBibitem
\bibitem[Hellmann(2022)]{Hellmann2022}
Hellmann,~R.
  \${{Ab}}{\textbackslash}phantom\{{\textbackslash}rule\{4pt\}\{0ex\}\}initio\$
  Determination of the Polarizability of Neon. \emph{Phys. Rev. A}
  \textbf{2022}, \emph{105}, 022809\relax
\mciteBstWouldAddEndPuncttrue
\mciteSetBstMidEndSepPunct{\mcitedefaultmidpunct}
{\mcitedefaultendpunct}{\mcitedefaultseppunct}\relax
\EndOfBibitem
\bibitem[Mori \latin{et~al.}(2023)Mori, Scarlett, Bray, and Fursa]{Mori2023}
Mori,~N.~A.; Scarlett,~L.~H.; Bray,~I.; Fursa,~D.~V. Convergent Close-Coupling
  Calculations of Positron Scattering from Atomic Carbon. \emph{Phys. Rev. A}
  \textbf{2023}, \emph{107}, 032817\relax
\mciteBstWouldAddEndPuncttrue
\mciteSetBstMidEndSepPunct{\mcitedefaultmidpunct}
{\mcitedefaultendpunct}{\mcitedefaultseppunct}\relax
\EndOfBibitem
\bibitem[Newell and Baird(1965)Newell, and Baird]{Newell1965}
Newell,~A.~C.; Baird,~R.~C. Absolute {{Determination}} of {{Refractive
  Indices}} of {{Gases}} at 47.7 {{Gigahertz}}. \emph{Journal of Applied
  Physics} \textbf{1965}, \emph{36}, 3751--3759\relax
\mciteBstWouldAddEndPuncttrue
\mciteSetBstMidEndSepPunct{\mcitedefaultmidpunct}
{\mcitedefaultendpunct}{\mcitedefaultseppunct}\relax
\EndOfBibitem
\bibitem[Orcutt and Cole(1967)Orcutt, and Cole]{Orcutt1967}
Orcutt,~R.~H.; Cole,~R.~H. Dielectric {{Constants}} of {{Imperfect Gases}}.
  {{III}}. {{Atomic Gases}}, {{Hydrogen}}, and {{Nitrogen}}. \emph{The Journal
  of Chemical Physics} \textbf{1967}, \emph{46}, 697--702\relax
\mciteBstWouldAddEndPuncttrue
\mciteSetBstMidEndSepPunct{\mcitedefaultmidpunct}
{\mcitedefaultendpunct}{\mcitedefaultseppunct}\relax
\EndOfBibitem
\bibitem[Hohm and Thakkar(2012)Hohm, and Thakkar]{Hohm2012}
Hohm,~U.; Thakkar,~A.~J. New {{Relationships Connecting}} the {{Dipole
  Polarizability}}, {{Radius}}, and {{Second Ionization Potential}} for
  {{Atoms}}. \emph{J. Phys. Chem. A} \textbf{2012}, \emph{116}, 697--703\relax
\mciteBstWouldAddEndPuncttrue
\mciteSetBstMidEndSepPunct{\mcitedefaultmidpunct}
{\mcitedefaultendpunct}{\mcitedefaultseppunct}\relax
\EndOfBibitem
\bibitem[Lupinetti and Thakkar(2005)Lupinetti, and Thakkar]{Lupinetti2005}
Lupinetti,~C.; Thakkar,~A.~J. Polarizabilities and Hyperpolarizabilities for
  the Atoms {{Al}}, {{Si}}, {{P}}, {{S}}, {{Cl}}, and {{Ar}}: {{Coupled}}
  Cluster Calculations. \emph{The Journal of Chemical Physics} \textbf{2005},
  \emph{122}, 044301\relax
\mciteBstWouldAddEndPuncttrue
\mciteSetBstMidEndSepPunct{\mcitedefaultmidpunct}
{\mcitedefaultendpunct}{\mcitedefaultseppunct}\relax
\EndOfBibitem
\bibitem[Lesiuk and Jeziorski(2023)Lesiuk, and Jeziorski]{Lesiuk2023}
Lesiuk,~M.; Jeziorski,~B. First-Principles Calculation of the
  Frequency-Dependent Dipole Polarizability of Argon. \emph{Phys. Rev. A}
  \textbf{2023}, \emph{107}, 042805\relax
\mciteBstWouldAddEndPuncttrue
\mciteSetBstMidEndSepPunct{\mcitedefaultmidpunct}
{\mcitedefaultendpunct}{\mcitedefaultseppunct}\relax
\EndOfBibitem
\end{mcitethebibliography}
\end{document}